\documentclass[journal=jacsat,manuscript=article]{achemso}
\usepackage[version=3]{mhchem}
\usepackage{graphicx}
\usepackage{dcolumn}
\usepackage{bm}
\usepackage{bbm}
\usepackage{graphicx,amsmath}
\usepackage{tikz}
\usepackage{mhchem}
\usepackage{amsmath, amsthm}

\usetikzlibrary{quantikz}
\usepackage{subfigure}
\usepackage{algpseudocode}
\usepackage{algorithm}

\title{Quantum-assisted variational Monte Carlo}

\author{Longfei Chang}
\author{Zhendong Li}\email{zhendongli@bnu.edu.cn}
\author{Wei-Hai Fang}
\affiliation{Key Laboratory of Theoretical and Computational Photochemistry, Ministry of Education, College of Chemistry, Beijing Normal University, Beijing 100875, China}

\begin{document}

\begin{tocentry}
\includegraphics[width=8cm,keepaspectratio]{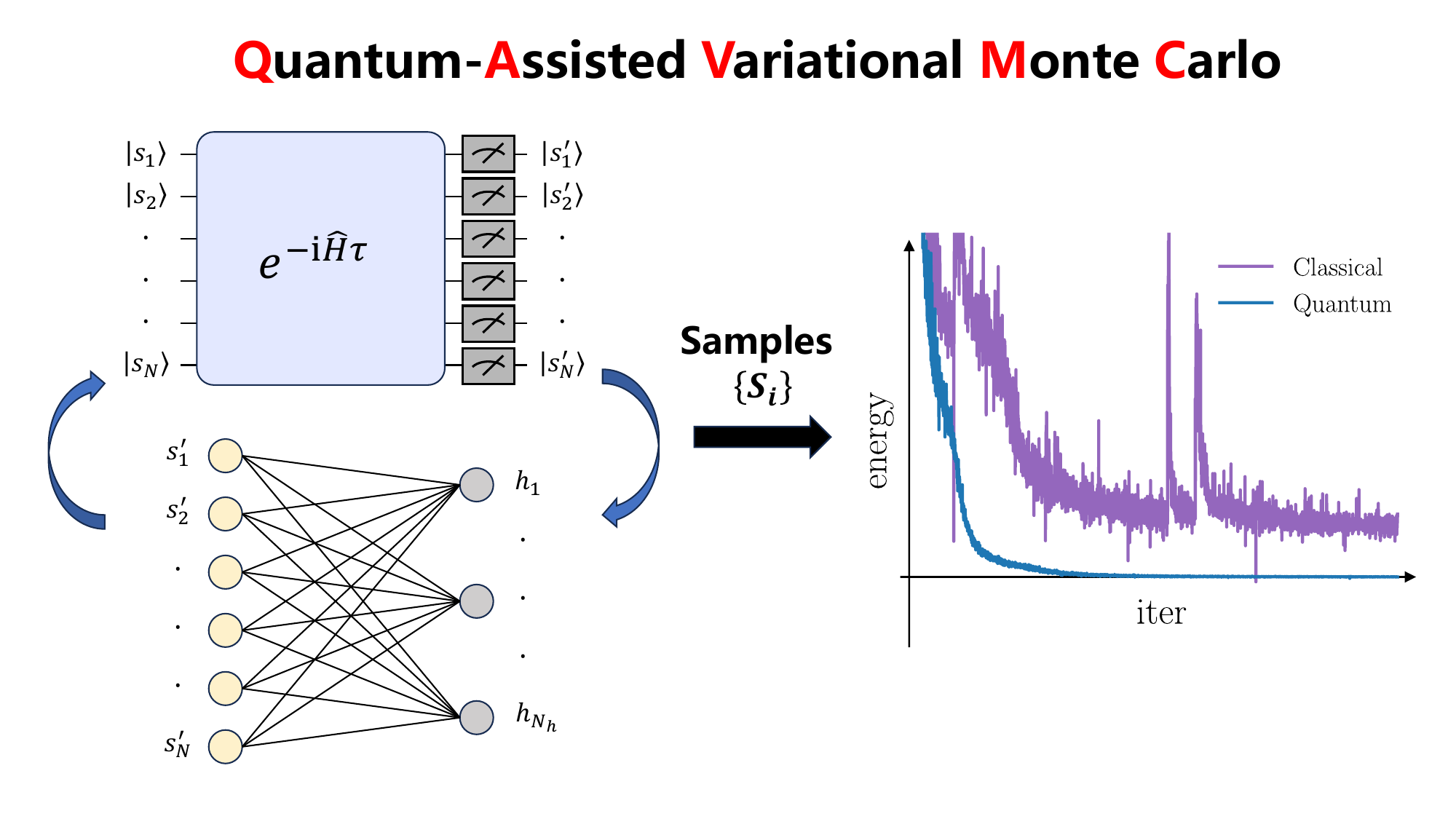}  





\end{tocentry}

\begin{abstract}
Solving the ground state of quantum many-body systems remains a fundamental challenge in physics and chemistry.
Recent advancements in quantum hardware have opened new avenues for addressing this challenge.
Inspired by the quantum-enhanced Markov chain Monte Carlo (QeMCMC) algorithm [Nature, 619, 282-287 (2023)], which was originally designed for sampling the Boltzmann distribution of classical spin models using quantum computers, we introduce a quantum-assisted variational Monte Carlo (QA-VMC) algorithm 
for solving the ground state of quantum many-body systems 
by adapting QeMCMC to sample the distribution of a (neural-network) wave function in VMC. 
The central question is whether such quantum-assisted proposal 
can potentially offer a computational advantage over classical methods.
Through numerical investigations for the Fermi-Hubbard model and molecular systems, we
demonstrate that the quantum-assisted proposal exhibits larger absolute spectral gaps and reduced autocorrelation times compared to conventional classical proposals, leading to more efficient sampling and faster convergence to the ground state in VMC as well as more accurate and precise estimation of physical observables.
This advantage is especially pronounced for specific parameter ranges, where the ground-state configurations are more concentrated in some configurations separated by large Hamming distances.
Our results underscore the potential of quantum-assisted algorithms to enhance classical variational methods for solving the ground state of quantum many-body systems.
\end{abstract}

\maketitle

\section{Introduction}
Accurately and efficiently solving the Schr\"odinger equation continues to pose a great challenge in quantum chemistry and condensed matter physics\cite{martin_interacting_2016}, primarily due to
the exponential growth of the Hilbert space with increasing system size.
To address this fundamental issue, a variety of classical computational methods have been developed, including density functional theory\cite{hohenberg_inhomogeneous_1964,kohn_self-consistent_1965,runge_density-functional_1984} (DFT), coupled cluster theory\cite{cizek_correlation_1966, purvis_full_1982,crawford_introduction_2000,shavitt_many-body_2009} (CC), density matrix renormalization group\cite{white_density_1992,chan_density_2011} (DMRG), various quantum Monte Carlo \cite{ceperley_ground_1980,zhang_constrained_1997,booth_fermion_2009,carleo_solving_2017,hermann_deep-neural-network_2020} (QMC) algorithms. Among these, variational Monte Carlo\cite{mcmillan1965ground} (VMC) has attracted significant attention in the era of artificial intelligence\cite{carleo_solving_2017,hermann_deep-neural-network_2020}, particularly
as neural networks (NNs) have emerged as a promising class of variational wave functions. Carleo and Troyer\cite{carleo_solving_2017} first employed the restricted Boltzmann machines (RBM),
a class of powerful energy-based models widely employed in machine learning for approximating discrete probability distributions\cite{le_roux_representational_2008},
as variational ansatz for spin systems and achieved high accuracy comparable to that of tensor network methods. 
This work has inspired subsequent research employing other
machine learning models, such as convolutional neural networks (CNNs)\cite{yang_deep_2020,wang_variational_2024},
autoregressive models\cite{hibat-allah_recurrent_2020,barrett_autoregressive_2022,wu_tensor-network_2023},
and Transformers\cite{wu2023nnqs,viteritti_transformer_2023,cao2024visiontransformerneuralquantum},
to tackle quantum many-body problems formulated in the framework of second quantization. 
For related studies addressing the solution of the Schr\"{o}dinger equation using NNs within the first quantization framework, we refer the reader to Ref. \cite{hermann2023ab} and the references cited therein. These advancements highlight the growing synergy between VMC
and machine learning, offering new avenues for solving complex quantum systems in physics and chemistry.

A key step in VMC is sampling configurations from the probability distribution of trial wave functions. 
The Markov chain Monte Carlo (MCMC) algorithm is one of the most widely used methods for this purpose\cite{levin_markov_2017}. 
However, it may face difficulties, such as prolonged mixing times, for challenging situations. 
For instance, in classical systems at critical points, the critical slowing down\cite{wolff1990critical} 
can significantly increase the mixing time of the Markov chain, making sampling inefficient.
Similar problems may also happen in sampling the ground-state distribution of quantum systems, 
such that a larger number of samples is required to achieve accurate energy estimates, thereby reducing the overall efficiency of the VMC algorithm\cite{jiang2024walkinghilbertspacequantum}.
To address these limitations, autoregressive neural networks have emerged as a promising alternative. By parameterizing the electronic wave function using autoregressive architectures \cite{hibat-allah_recurrent_2020,barrett_autoregressive_2022,wu_tensor-network_2023}, efficient and scalable sampling 
based on conditional distribution can be achieved without relying on MCMC. However, many previously mentioned NN wave functions without such autoregressive structure, including RBM, CNNs, and vision transformers \cite{viteritti_transformer_2023,cao2024visiontransformerneuralquantum} (ViTs),
still rely on MCMC for sampling. 
Therefore, there persists a critical need for developing innovative strategies to enhance sampling efficiency in VMC.

Thanks to the rapid development of quantum hardware\cite{arute_quantum_2019,wu_strong_2021}, quantum computation has become a promising tool for tackling challenging computational problems\cite{cao_quantum_2019,mcardle_quantum_2020,bauer_quantum_2020,motta_emerging_2022}. Many quantum algorithms have been proposed to accelerate sampling from the Gibbs state or the classical Boltzman distribution \cite{szegedy_quantum_2004,somma_quantum_2008,wocjan_speedup_2008,poulin_sampling_2009,bilgin_preparing_2010,temme_quantum_2011,yung_quantumquantum_2012,montanaro_quantum_2015,chowdhury_quantum_2017,lemieux_efficient_2020,arunachalam_simpler_2022,rall_thermal_2023,chen_efficient_2023,wild_quantum_2021,wild_quantum_2021-1,layden_quantum-enhanced_2023,nakano_markov-chain_2024,PhysRevResearch.6.033147}. In particular, the recently proposed quantum-enhanced Markov chain Monte Carlo (QeMCMC) algorithm\cite{layden_quantum-enhanced_2023} stands out as a hybrid quantum-classical method for sampling from the Boltzmann distribution of classical spin systems, which has been shown to accelerate the convergence of Markov chain for spin-glass models at low temperatures both numerically and
experimentally on near-term quantum devices \cite{layden_quantum-enhanced_2023}. This work has spurred
several further developments\cite{orfi_bounding_2024, orfi_barriers_2024,nakano_markov-chain_2024,christmann_quantum_2024,lockwood_quantum_2024,ferguson_quantum-enhanced_2024,sajjan2024polynomiallyefficientquantumenabled}, including investigations into the limitations of the algorithm \cite{orfi_bounding_2024, orfi_barriers_2024}, the use of quantum alternating operator ansatz as an alternative to time evolution
to reduce circuit depth\cite{nakano_markov-chain_2024}, and the development of quantum-inspired sampling algorithms based on QeMCMC\cite{christmann_quantum_2024}, and improving sampling efficiency of VMC through surrogate models\cite{sajjan2024polynomiallyefficientquantumenabled}.

In this work, inspired by the QeMCMC algorithm \cite{layden_quantum-enhanced_2023} for sampling classical Boltzmann distributions, we propose a quantum-assisted VMC (QA-VMC) algorithm to address the sampling challenges for solving quantum many-body problems using VMC. Similar to QeMCMC, our approach leverages the unique capability of quantum computers to perform time evolution and utilizes the resulting quantum states to propose new configurations, while all other components of the algorithm are executed on classical computers to minimize the demand for quantum resources.
A central question we aim to explore in this work is whether QA-VMC can offer a potential advantage in sampling the ground state distributions of quantum many-body systems. To investigate this, we benchmark the algorithm against classical 
sampling methods for various models,
including the Fermi-Hubbard model (FHM) and molecular systems,
with different system sizes and parameters. The remainder of this paper is structured as follows. First, we provide a concise overview of the VMC algorithm and MCMC sampling techniques. Next, we introduce the QA-VMC algorithm and the figures of merit used to evaluate the convergence of different MCMC algorithms. Subsequently, we present the results of the quantum-assisted algorithm for various systems and compare its performance against classical sampling methods. Finally, we summarize our findings and discuss future directions.

\section{Theory and algorithms}
\subsection{Variational Monte Carlo}
The VMC\cite{sorella_wave_2005, pfau_ab_2020, choo_fermionic_2020} method is a computational algorithm that combines the variational principle with Monte Carlo sampling to approximate the ground state of a Hamiltonian $\hat{H}$ using a trial wave function. 
Specifically, for a variational
wave function $\lvert\psi_{\boldsymbol{\theta}}\rangle$ characterized by  a set of variational parameters $\boldsymbol{\theta}$,
the energy function can be expressed as 
\begin{eqnarray}
E_{\boldsymbol{\theta}}=\frac{\langle\psi_{\boldsymbol{\theta}}\lvert\hat{H}\lvert\psi_{\boldsymbol{\theta}}\rangle}{\langle\psi_{\boldsymbol{\theta}}\lvert\psi_{\boldsymbol{\theta}}\rangle}=\sum_{\boldsymbol{S}} P_{\boldsymbol{\theta}}(\boldsymbol{S})E^{\mathrm{loc}}_{\boldsymbol{\theta}}(\boldsymbol{S}),
\end{eqnarray}
where the configuration $\boldsymbol{S}\equiv(s_1,\cdots,s_N)$ consists of spins (or qubits) $s_j=\pm 1$. The probability distribution is defined as
$P_{\boldsymbol{\theta}}(\boldsymbol{S})\equiv\lvert\langle\boldsymbol{S}\lvert\psi_{\boldsymbol{\theta}}\rangle\lvert^2/\langle\psi_{\boldsymbol{\theta}}\lvert\psi_{\boldsymbol{\theta}}\rangle$, and
the local energy is given by
\begin{eqnarray}
    E^{\mathrm{loc}}_{\boldsymbol{\theta}}(\boldsymbol{S})\equiv\frac{\bra{\boldsymbol{S}}\hat{H}\ket{\psi_{\boldsymbol{\theta}}}}{\langle\boldsymbol{S}\lvert\psi_{\boldsymbol{\theta}}\rangle}=\frac{\sum_{\boldsymbol{S}'}\langle\boldsymbol{S}\lvert\hat{H}\lvert\boldsymbol{S}'\rangle\langle\boldsymbol{S}'\lvert\psi_{\boldsymbol{\theta}}\rangle}{\langle\boldsymbol{S}\lvert\psi_{\boldsymbol{\theta}}\rangle}.\label{eq:elocal}
\end{eqnarray}
In the VMC framework, the energy function is approximated using the Monte Carlo algorithm by sampling configurations $\{\boldsymbol{S}_i\}$ from $P_{\boldsymbol{\theta}}(\boldsymbol{S})$, i.e.,
\begin{eqnarray}
    E_{\boldsymbol{\theta}}\approx\frac{1}{N_s}\sum_{i=1}^{N_s} E^{\mathrm{loc}}_{\boldsymbol{\theta}}(\boldsymbol{S}_i),
\end{eqnarray}
where $N_s$ denotes the number of samples. Similarly, the energy gradients with respect to the parameters can be estimated
as\cite{sorella_wave_2005}
\begin{eqnarray}
    \frac{\partial E_{\boldsymbol{\theta}}}{\partial \boldsymbol{\theta}}\approx\frac{1}{N_s}\sum_{i=1}^{N_s}2\Re{\Big[\left(E_{\boldsymbol{\theta}}^{\mathrm{loc}}(\boldsymbol{S}_i)-E_{\boldsymbol{\theta}}\right)\frac{\partial \ln{\psi_{\boldsymbol{\theta}}^{*}(\boldsymbol{S}_i)}}{\partial \boldsymbol{\theta}}\Big]}.
\end{eqnarray}
For sparse Hamiltonians, the local energy \eqref{eq:elocal} can be computed with polynomial
cost with respect to the system size $N$, provided that the value of trial wave function $\psi_{\boldsymbol{\theta}}(\boldsymbol{S}_i)$ can be evaluated with polynomial cost. 
Consequently, VMC enables efficient estimation of the energy 
and optimization of the parameters,
even for highly complex wave function ans\"{a}tze for which the overlap
$\langle\psi_{\boldsymbol{\theta}}|\psi_{\boldsymbol{\theta}}\rangle$
and the expectation value of the Hamiltonian 
$\langle\psi_{\boldsymbol{\theta}}|\hat{H}|\psi_{\boldsymbol{\theta}}\rangle$
cannot be efficiently computed exactly.

The accuracy of VMC calculations is strongly dependent on the flexibility of the wave function ansatz. The RBM ansatz\cite{carleo_solving_2017} 
for the wave function $\lvert \psi_{\boldsymbol{\theta}}\rangle = \sum_{\boldsymbol{S}}\psi_{\boldsymbol{\theta}}(\boldsymbol{S})\lvert \boldsymbol{S}\rangle$ can be expressed as
\begin{gather}
    \psi_{\boldsymbol{\theta}}(\boldsymbol{S})=\sum_{\boldsymbol{h}}\exp{\big(E^{\mathrm{RBM}}_{\boldsymbol{\theta}}(\boldsymbol{S})\big)}, \\
    E^{\mathrm{RBM}}_{\boldsymbol{\theta}}(\boldsymbol{S})=\sum_{i=1}^N a_i s_i+\sum_{\mu=1}^M b_{\mu} h_{\mu} +\sum_{i=1}^N\sum_{\mu=1}^M s_i W_{\mu i}h_{\mu},
\end{gather}
where $h_{\mu}\in\{-1,1\}$ is a set of binary hidden variables, and the set of real or complex variables
$\boldsymbol{\theta}=\{W_{\mu i}, a_i, b_\mu\}$ are variational parameters. Here, $W_{\mu i}$ denotes the weights connecting variables $s_i$ and $h_\mu$, and $a_i$ and $b_\mu$ 
are the biases associated with the physical variables $s_i$ and hidden variables $h_\mu$,
respectively. The representational power of RBM increases with the number of hidden variables $M$, and the density of 
hidden units, defined as $\alpha\equiv M/N$, serves as a measure of the model's complexity.
In this work, we utilized the RBMmodPhase ansatz\cite{torlai_neural-network_2018} implemented in
the NetKet package\cite{choo_fermionic_2020} as trial wave functions.
This ansatz employs two RBMs
with real parameters,
denoted by $A_{\boldsymbol{\theta}}(\boldsymbol{S})$ and
$B_{\boldsymbol{\phi}}(\boldsymbol{S})$,
to separately model the amplitude and phase of the wave function, i.e.,
$\psi_{\boldsymbol{\theta,\phi}}(\boldsymbol{S}) = A_{\boldsymbol{\theta}}(\boldsymbol{S}) e^{\mathbbm{i}\ln{B_{\boldsymbol{\phi}}(\boldsymbol{S})}}$. For optimization, we employed the stochastic reconfiguration method\cite{sorella_wave_2005} in conjunction with the Adam optimizer\cite{kingma2017adammethodstochasticoptimization}.

\subsection{Markov chain Monte Carlo}
To sample configurations from the probability distribution $P_{\boldsymbol{\theta}}(\boldsymbol{S})$, the MCMC algorithm is commonly employed in VMC.
MCMC generates samples from a target probability distribution $\pi(\boldsymbol{S})$ by constructing a Markov chain that explores a defined state space $\{\boldsymbol{S}_i\}$.
The transition from state $\boldsymbol{S}_i$ to state $\boldsymbol{S}_j$ 
is governed by a transition probability $\mathcal{P}(\boldsymbol{S}_i,\boldsymbol{S}_j)$.
If the Markov chain is \emph{irreducible} and \emph{aperiodic}, 
it is guaranteed to converge to a unique stationary
distribution\cite{levin_markov_2017}, which corresponds to the target distribution $\pi(\boldsymbol{S})$.
A sufficient condition to ensure this convergence is the detailed balance condition expressed as
\begin{eqnarray}
\pi(\boldsymbol{S}_i)\mathcal{P}(\boldsymbol{S}_i,\boldsymbol{S}_j)=\pi(\boldsymbol{S}_j)\mathcal{P}(\boldsymbol{S}_j,\boldsymbol{S}_i),\quad\forall i,j.
\end{eqnarray}
One of the most widely used sampling methods that satisfies the detailed balance condition is the Metropolis-Hastings algorithm\cite{metropolis_equation_1953}.
This algorithm decomposes the transition process into two steps: first, a candidate move is proposed according to a proposal distribution
$\mathcal{Q}(\boldsymbol{S}_i,\boldsymbol{S}_{j})$, and second, the move is either accepted or rejected based on an acceptance probability 
$\mathcal{A}(\boldsymbol{S}_i,\boldsymbol{S}_j)$,
defined as
\begin{eqnarray}
\mathcal{A}(\boldsymbol{S}_i,\boldsymbol{S}_j)=\mathrm{min}\big(1,\frac{\pi(\boldsymbol{S}_{j})\mathcal{Q}(\boldsymbol{S}_{j},\boldsymbol{S}_{i})}{\pi(\boldsymbol{S}_{i})\mathcal{Q}(\boldsymbol{S}_{i},\boldsymbol{S}{j})}\big).\label{eq: acceptance}
\end{eqnarray}
Using this approach, a Markov chain can be constructed for any target probability distribution
$\pi(\boldsymbol{S})$ on the state space $\{\boldsymbol{S}_i\}$, 
with a transition matrix $\mathcal{P}$ given by
\begin{eqnarray}
\mathcal{P}(\boldsymbol{S}_i, \boldsymbol{S}_j) =
\begin{cases} 
\mathcal{Q}(\boldsymbol{S}_i, \boldsymbol{S}_j)\mathcal{A}(\boldsymbol{S}_i, \boldsymbol{S}_j) & \text{if } \boldsymbol{S}_j\neq\boldsymbol{S}_i,\\
1 - \sum_{\boldsymbol{S}' \neq \boldsymbol{S}_i} \mathcal{Q}(\boldsymbol{S}_i, \boldsymbol{S}') \mathcal{A}(\boldsymbol{S}_i, \boldsymbol{S}') & \text{if } \boldsymbol{S}_j=\boldsymbol{S}_i.
\end{cases}\label{eq:transitionMatrix}
\end{eqnarray}
The proposal distribution $\mathcal{Q}(\boldsymbol{S}_{i},\boldsymbol{S}_{j})$
can take nearly any form, provided it is efficiently computable.
However, since different $\mathcal{Q}(\boldsymbol{S}_{i},\boldsymbol{S}_{j})$ will
result in different $\mathcal{P}(\boldsymbol{S}_{i},\boldsymbol{S}_{j})$,
its choice has a significant impact on the convergence rate of the MCMC algorithm. 
A well-designed proposal distribution can significantly enhance sampling efficiency, enabling faster exploration of the state space.
On the other hand, a poorly chosen proposal distribution may result in slow convergence or inefficient exploration of the state space. 
For Fermionic systems, such as the FHMs and molecular systems, 
commonly employed proposals encompass the Uniform proposal (selecting a random configuration), the Exchange proposal (swapping occupations of two same-spin orbitals randomly), and the ExcitationSD proposal, which generates new configurations through restricted random excitations, similar to the Uniform proposal but limited to singles and double excitations.

Recently, Layden et al.\cite{layden_quantum-enhanced_2023} introduced 
the QeMCMC algorithm for sampling from the Boltzmann distribution $\pi(\boldsymbol{S})=\frac{1}{Z}e^{-\frac{E(\boldsymbol{S})}{T}}$ of the 'spin glass' Ising model, where the energy of a configuration $\boldsymbol{S}$ is given by
$E(\boldsymbol{S})=-\sum_{j>k=1}^n J_{jk} s_js_k-\sum_{j=1}^n h_j s_j$, with $T$ being the temperature
and $Z$ being the partition function.
In this approach, proposals are generated with the help of time evolution on quantum computers. Specifically, the time evolution operator $\hat{U}(\gamma,\tau)=\exp{(-\mathbbm{i}\hat{H}(\gamma)\tau)}$ is constructed from a specially designed
Hamiltonian
\begin{eqnarray}
    \hat{H}(\gamma)&=&(1-\gamma)\alpha\hat{H}_{\mathrm{prob}}+\gamma\hat{H}_{\mathrm{mix}},\label{eq:HamQeMCMC}
\end{eqnarray}
where $\hat{H}_{\mathrm{prob}}$ shares the same parameters with the problem and
$\hat{H}_{\mathrm{mix}}$ is a mixing term
\begin{eqnarray}    
    \hat{H}_{\mathrm{prob}}&=&-\sum_{j>k=1}^n J_{jk} \hat{Z}_j\hat{Z}_k-\sum_{j=1}^n h_j \hat{Z}_j,\\
    \hat{H}_{\mathrm{mix}}&=&\sum_{j=1}^n \hat{X}_j.
\end{eqnarray}
Here, $\alpha=\|\hat{H}_{\mathrm{mix}}\|_{\mathrm{F}}/\|\hat{H}_{\mathrm{prob}}\|_{\mathrm{F}}$ is a normalizing factor, and $\gamma\in [0,1]$ controls the relative weights of the two terms. The quantum proposal distribution is then defined as
\begin{eqnarray}
    \mathcal{Q}(\boldsymbol{S}_i, \boldsymbol{S}_j;\gamma, \tau)
    =
    \lvert \langle \boldsymbol{S}_j \lvert \exp{(-\mathbbm{i}\hat{H}(\gamma)\tau)} \lvert \boldsymbol{S}_i \rangle \lvert^2.
\end{eqnarray}
In the QeMCMC procedure\cite{layden_quantum-enhanced_2023}, $\gamma$ and $\tau$ are randomly selected within predefined ranges at each MCMC
step. Notably, since $\hat{H}=\hat{H}^T$ in Eq. \eqref{eq:HamQeMCMC}, 
it follows that  $\hat{U}=\hat{U}^T$ and $\mathcal{Q}=\mathcal{Q}^T$. 
Consequently, the acceptance probability in Eq. \eqref{eq: acceptance} simplifies to
\begin{eqnarray}
\mathcal{A}(\boldsymbol{S}_i,\boldsymbol{S}_j)=\mathrm{min}\big(1,\frac{\pi(\boldsymbol{S}_{j})}{\pi(\boldsymbol{S}_{i})}\big),\label{eq:Aqemc}
\end{eqnarray} 
which avoids the explicit computation of $\mathcal{Q}$.
Numerical and experimental results demonstrate that this quantum proposal leads to faster convergence at low temperatures compared to classical local and uniform moves\cite{layden_quantum-enhanced_2023}. 
This improvement is attributed to the ability of the quantum proposal to generate moves that result in small energy changes $|\Delta E|=|E(\boldsymbol{S}_i)
-E(\boldsymbol{S}_j)|$, while achieving large Hamming distances, thus enhancing exploration efficiency
for challenging distributions.

\subsection{Quantum-assisted variational Monte Carlo}
\begin{figure}[htbp]
    \centering
    \includegraphics[width=0.7\textwidth]{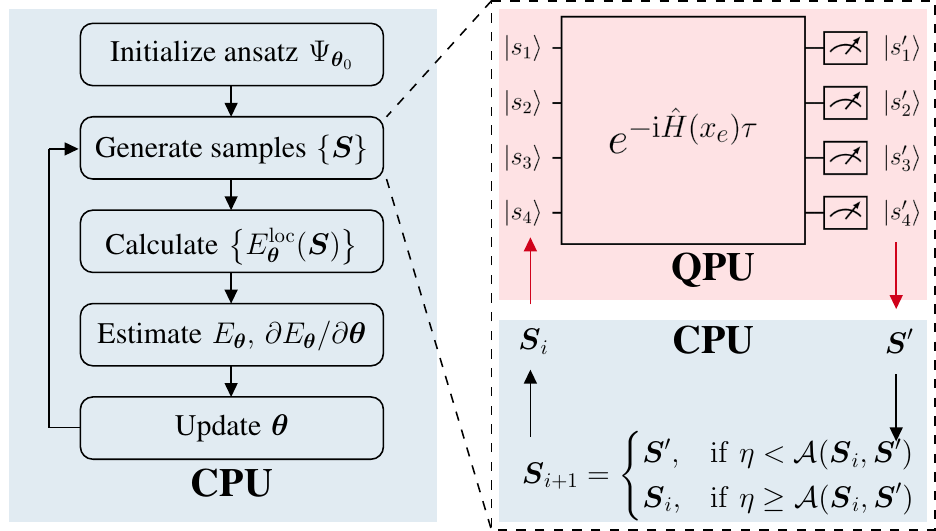}
    \caption{Flowchart of the QA-VMC algorithm. The red box highlights the quantum step executed
    on quantum processor units (QPU), where
    a quantum time-evolution governed by a chosen Hamiltonian $\hat{H}$ 
    satisfying $\hat{H}^T=\hat{H}$ and measurements are employed to propose new configurations.   
    All other parts of the algorithm are executed on classical computers.
    The acceptance probability
    $\mathcal{A}(\boldsymbol{S}_i,\boldsymbol{S}_j)$ is determined by Eq. \eqref{eq:Aqemc}, and
    $\eta\in[0,1]$ is a uniformly distributed random number.
    }
    \label{fig: QA_VMC}
\end{figure}

Inspired by the QeMCMC algorithm \cite{layden_quantum-enhanced_2023} for sampling classical Boltzmann distributions, we propose the QA-VMC algorithm, as illustrated in Figure \ref{fig: QA_VMC}, for solving quantum many-body systems. The code is available in Ref.
\cite{changQAVMC2024}.
Given a problem specified by the Hamiltonian $\hat{H}(x)$, which depends on a parameter $x$ such as the on-site interaction $U$ in FHM, we propose generating
new configurations using the time evolution operator $\hat{U}(x_e,\tau)=\exp(-\mathbbm{i}\hat{H}(x_e)\tau)$, where $x_e$ may differ from $x$ to optimize sampling efficiency. For real Hamiltonians considered in this work, the Hermiticity of $\hat{H}$ ensures that it is also symmetric, such that Eq. \eqref{eq:Aqemc} still holds.
We will refer to this proposal as the Quantum proposal in the following context.
Very recently, Ref. \cite{sajjan2024polynomiallyefficientquantumenabled} proposed another way to combine VMC and QeMCMC, where a surrogate network based on the classical Ising model is introduced to first fit the target distribution. The QeMCMC\cite{layden_quantum-enhanced_2023} algorithm is then directly applied to sample the probability distribution of the surrogate network, and the energy in VMC is estimated using reweighting technique. In contrast, our approach is much simpler, since it does not require the fitting procedure. However, in our case, a good effective Hamiltonian $\hat{H}(x_e)$ driving the unitary evolution needs to be designed for efficient sampling.

To gain a deeper understanding of the Quantum proposal, we decompose the corresponding proposal probability
$\mathcal{Q}^{\mathrm{q}}(\boldsymbol{S}_i, \boldsymbol{S}_j;x_e,\tau)$
into two parts
\begin{eqnarray}
    \mathcal{Q}^{\mathrm{q}}(\boldsymbol{S}_i, \boldsymbol{S}_j;x_e,\tau)&=&\lvert \langle \boldsymbol{S}_j \lvert \exp(-\mathbbm{i}\hat{H}(x_e)\tau) \lvert \boldsymbol{S}_i \rangle \lvert^2\nonumber\\
    &=&\sum_n p_{n}(\boldsymbol{S}_i;x_e)p_{n}(\boldsymbol{S}_j;x_e)+\Omega(\boldsymbol{S}_i,\boldsymbol{S}_j;x_e,\tau)\label{eq: Quantum_proposal},
\end{eqnarray}
where $p_{n}(\boldsymbol{S}_i;x_e) = \lvert \langle \boldsymbol{S}_i \lvert \Psi_n \rangle \lvert^2$ and 
$\{\lvert \Psi_n \rangle\}$ represents the eigenstates of $\hat{H}(x_e)$, and 
$\Omega(\boldsymbol{S}_i,\boldsymbol{S}_j;x_e,\tau)$ is given by
\begin{eqnarray}
\Omega(\boldsymbol{S}_i,\boldsymbol{S}_j;x_e,\tau)=2\Re{\sum_{n>m} \langle \boldsymbol{S}_j \lvert \Psi_n \rangle \langle \Psi_n \lvert \boldsymbol{S}_i \rangle \langle \boldsymbol{S}_i \lvert \Psi_m \rangle \langle \Psi_m \lvert \boldsymbol{S}_j \rangle e^{\mathbbm{i}(E_m-E_n)\tau}}.\label{eq: time_depend_term}
\end{eqnarray}
The first term in Eq. \eqref{eq: Quantum_proposal} is time-independent and will be referred to as the Effective proposal
\begin{eqnarray}
    \mathcal{Q}^{\mathrm{eff}}(\boldsymbol{S}_i, \boldsymbol{S}_j;x_e) = \sum_n 
    p_{n}(\boldsymbol{S}_i;x_e)p_{n}(\boldsymbol{S}_j;x_e),\label{eq: Effective_proposal}
\end{eqnarray}
since it can be verified that $\sum_{\boldsymbol{S}_j} \mathcal{Q}^{\mathrm{eff}}(\boldsymbol{S}_i, \boldsymbol{S}_j;x_e)=1$. While
$\mathcal{Q}^{\mathrm{eff}}(\boldsymbol{S}_i, \boldsymbol{S}_j;x_e)$ is inefficient to implement on classical computers 
and quantum computers directly, it provides valuable insights into the usefulness of the Quantum proposal 
based on the following observations:

First, for a Hamiltonian $\hat{H}(x_e)$ without degeneracy, the time-averaged $\mathcal{Q}^{\mathrm{q}}$ over $\tau\in (-\infty, +\infty)$ equals $\mathcal{Q}^{\mathrm{eff}}$, i.e.,
\begin{eqnarray}
    \mathcal{Q}^{\mathrm{eff}}(\boldsymbol{S}_i,\boldsymbol{S}_j;x_e)=\lim_{\tau\rightarrow+\infty}\frac{1}{2\tau}\int_{-\tau}^{+\tau}\mathcal{Q}^{\mathrm{q}}(\boldsymbol{S}_i,\boldsymbol{S}_j;x_e,\tau')\mathrm{d}\tau'.\label{eq: relation}
\end{eqnarray}
This implies that if we randomly select $\tau$ within some sufficiently large interval $(-T,+T)$, the averaged $\mathcal{Q}^{\mathrm{q}}$ will equal $\mathcal{Q}^{\mathrm{eff}}$. This point is further illustrated
in Supporting Information for different model systems.

Second, the proposed move using $\mathcal{Q}^{\mathrm{eff}}$ has a more intuitive interpretation, because Eq. \eqref{eq: Effective_proposal} can be understood
as follows: given a configuration $\boldsymbol{S}_i$, first randomly select an eigenstate $\ket{\Psi_n}$ according to the conditional probability distribution $P(n|\boldsymbol{S}_i)\equiv p_n(\boldsymbol{S}_i;x_e)$,
and then randomly select a configuration $\boldsymbol{S}_j$
based on the conditional probability distribution $P(\boldsymbol{S}_j|n)\equiv p_n(\boldsymbol{S}_j;x_e)$.
Thus, if $p_0(\boldsymbol{S}_i)$ and $p_0(\boldsymbol{S}_j)$ for the ground state are both large, $Q^{\mathrm{eff}}(\boldsymbol{S}_i, \boldsymbol{S}_j;x_e)$ will also be large, regardless of the Hamming distance between $\boldsymbol{S}_i$ and $\boldsymbol{S}_j$. This suggests that for a ground state probability distribution concentrated on some configurations with large Hamming distances, the Effective proposal can offer a significant advantage over classical proposals. Based on Eq. \eqref{eq: relation}, we expect the Quantum proposal to exhibit similar behavior.

A primary objective of this work is to examine whether the QA-VMC algorithm can potentially enhance the convergence of MCMC simulations, thereby providing computational efficiency gains for VMC. 
To investigate this, we apply this algorithm to FHMs and molecular systems across various parameter ranges and system sizes. Through a comprehensive comparative analysis with conventional classical proposals, we evaluate the performance of QA-VMC from multiple perspectives, as detailed in the following section.

\subsection{Figures of merit}
\subsubsection{Absolute spectral gap}
The convergence rate of the Markov chain can be quantitatively characterized by its mixing time\cite{levin_markov_2017,layden_quantum-enhanced_2023} $t_{\mathrm{mix}}(\varepsilon)$,
which is the minimum number of steps $t$
required for the Markov chain to converge to its stationary distribution within a predefined tolerance threshold $\varepsilon$, i.e.,
\begin{eqnarray}
    t_{\mathrm{mix}}(\varepsilon):=\min \{t:\max_{\boldsymbol{S}_i}\lVert \mathcal{P}^t(\boldsymbol{S}_i,\cdot)-\pi(\cdot) \lVert_{\mathrm{TV}}\leq \varepsilon\}, \label{eq: mixing_time_def}
\end{eqnarray}
where $\lVert \cdot\lVert_{\mathrm{TV}}$ denotes the total variation distance\cite{levin_markov_2017}, quantifying the discrepancy between the chain's distribution after $t$ steps and the stationary distribution.
While the exact computation of $t_{\mathrm{mix}}(\varepsilon)$ is
generally intractable, it can be effectively bounded by the absolute spectral gap $\delta$ via\cite{levin_markov_2017}
\begin{eqnarray}
    (\delta^{-1}-1)\ln{\Big(\frac{1}{2\varepsilon}\Big)}\leq t_{\mathrm{mix}}(\varepsilon)\leq \delta^{-1} \ln{\Big(\frac{1}{\varepsilon \min_{\boldsymbol{S}}\pi(\boldsymbol{S})}\Big)}.\label{eq: mixing_time}
\end{eqnarray}
Here, $\delta = 1 - |\lambda_2|\in[0,1]$ is the difference between the absolute values of the two largest eigenvalues ($\lambda_1=1$ and $\lambda_2$) of the transition matrix $\mathcal{P}$ \eqref{eq:transitionMatrix},
which can be computed through matrix diagonalization, making $\delta$ more readily accessible than the mixing time.
As evident from Eq.\eqref{eq: mixing_time}, the spectral gap $\delta$ exhibits an inverse relationship with the bounds of the mixing time, thereby serving as a precise quantitative measure for assessing Markov chain convergence\cite{layden_quantum-enhanced_2023}. Specifically, a larger spectral gap $\delta$ implies 
smaller $t_{\mathrm{mix}}(\varepsilon)$ and hence faster convergence to the stationary distribution.
However, it is crucial to acknowledge that the practical computation of $\delta$ is
limited by the exponential growth of the Hilbert space.
Therefore, in this work we employ an extrapolation approach adopted
in the QeMCMC work\cite{layden_quantum-enhanced_2023} to establish a 
relationship between $\delta$
and system size $N$ obtained from computationally feasible systems. 
This enables us to estimate the asymptotic behavior of $\delta$
for larger systems that are infeasible for diagonalization.

\subsubsection{Autocorrelation time}
Apart from the absolute spectral gap, autocorrelation time is another valuable metric
for assessing the convergence of MCMC algorithms\cite{Sokal1997}.
This metric is widely used in practice because it directly captures the convergence behavior of the Markov chain, particularly in terms of how long the chain retains memory of its previous states. For a given operator $\hat{O}$, the integrated autocorrelation time $\tau_{O}$ is defined as
\begin{eqnarray}
    \tau_{O}= 1 + 2\sum_{\tau=1}^{\infty}\rho_{O}(\tau),\quad
    \rho_{O}(\tau)=\frac{c_{O}(\tau)}{c_{O}(0)},\label{eq:tauint}
\end{eqnarray}
where $c_{O}(\tau)$ represents the autocovariance function at lag $\tau$
\begin{eqnarray}
    c_{O}(\tau)=\frac{\sum_{i=1}^{N_s-\tau}(O_{\mathrm{loc}}(\boldsymbol{S}_i)-\mu_O)(O_{\mathrm{loc}}(\boldsymbol{S}_{i+\tau})-\mu_O)}{N_s-\tau}.
\end{eqnarray}
Here, $O_{\mathrm{loc}}(\boldsymbol{S}_i)\equiv\frac{\langle\boldsymbol{S}_i\lvert\hat{O}\lvert\Psi\rangle}{\langle\boldsymbol{S}_i\lvert\Psi\rangle}$, $\mu_O
=\frac{1}{N_s}\sum_{i=1}^{N_s}O_{\mathrm{loc}}(\boldsymbol{S}_i)$ denotes 
the sample average, and $N_s$ represents the sample size. 
A smaller $\tau_{O}$ indicates faster convergence
of the estimator to its mean, reflecting efficient mixing of the chain. 
Conversely, a larger $\tau_{O}$ suggests strong correlations among samples and slow mixing.
The integrated autocorrelation time is related to the effective sample size $N_{\mathrm{eff}}$ by $N_{\mathrm{eff}}=N_s/\tau_{O}$. Thus, it can serve as a practical and intuitive measure of the chain's convergence properties. 
We used the algorithm introduced in Ref. \cite{Sokal1997} to estimate $\tau_O$.

\subsubsection{Metric for potential quantum speedup}
To explore the potential quantum speedup of the Quantum proposal compared to classical proposals, 
we analyze the asymptotic behavior of the quantity $\mathcal{T}_{\mathrm{eff}}=\delta^{-1}t_{s}$,
which will be referred to as the effective runtime.
Here, $\delta^{-1}$ estimates the number of steps required to reach equilibrium,
and $t_s$ is the runtime of a single execution of a classical or quantum move.
Thus, $\mathcal{T}_{\mathrm{eff}}$ roughly estimates the runtime of an ideal MCMC algorithm.
The spectral gap $\delta$ can be modeled by 
an exponential function
with respect to the system size $N$ via $\delta(N) = a 2^{-kN}$\cite{layden_quantum-enhanced_2023}.
Then, the ratio between the effective runtime
of a classical proposal $\mathcal{T}_{\mathrm{eff},c}$ and that of the
Quantum proposal $\mathcal{T}_{\mathrm{eff},q}$ proposals can be expressed as
\begin{eqnarray}
\frac{\mathcal{T}_{\mathrm{eff},c}}{\mathcal{T}_{\mathrm{eff},q}}
=
\frac{\delta_c^{-1}t_{s,c}}{\delta_{q}^{-1}t_{s,q}}
=
\frac{a_{q}t_{s,c}}{a_{c}t_{s,q}}
2^{(k_c-k_q)N}.\label{eq:classicalVsQuantum}
\end{eqnarray}
The runtime $t_{s,c}$ for classical moves considered in this work scales at most polynomially
with the system size $N$. Consequently, if 
the runtime $t_{s,q}$ for the quantum case also scales polynomially, then 
$\mathcal{T}_{\mathrm{eff},c}>\mathcal{T}_{\mathrm{eff},q}$
for sufficiently large systems, provided that $k_c>k_q$. 
However, if $t_{s,q}$ scales exponentially as $O(2^{bN})$, 
a potential speedup can only exist if $k_c>k_q+b$.
Therefore, in addition to the asymptotic behavior of $\delta$
characterized by the exponent $k$,
the potential quantum advantage is also critically dependent on the 
scaling of $t_{s,q}$ with respect to $N$.
In the following sections, we will focus on the asymptotic behaviors of
both $\delta$ and $t_{s,q}$.

\section{Results and discussion}

\subsection{Fermi-Hubbard model}
We begin by evaluating the performance of the QA-VMC algorithm for the FHM\cite{arovas_hubbard_2022},
which serves as a benchmark for both classical and quantum variational methods \cite{yokoyama_variational_1987,cade_strategies_2020}.
The Hamiltonian of the FHM is given by:
\begin{gather}
    \hat{H}(U)=-t\sum_{\langle i,j \rangle}\sum_{\sigma}(\hat{a}_{i\sigma}^\dagger \hat{a}_{j\sigma}+\mathrm{h.c.})+U\sum_{i}\hat{n}_{i\alpha}\hat{n}_{i\beta},
\end{gather}
where the hopping parameter $t=1$, $U$ is the on-site interaction, $\sigma \in \{\alpha,\beta\}$, $\hat{a}_q^{(\dagger)}$ represent Fermionic annihilation (creation) operators, and 
$\langle i,j \rangle$ represents the summation over nearest-neighbor sites. Additionally, we use the Jordan-Wigner mapping\cite{jordan1928pauli} to transform the Fermionic Hamiltonian $\hat{H}$ into a qubit Hamiltonian expressed as a linear combination of Pauli terms, i.e. $\hat{H}=\sum_k h_k P_k$ with $P_k\in\{I,X,Y,Z\}^{\otimes N}$, and the occupation number vectors into corresponding qubit configurations.
In this study, we focus on the ground state of the FHM with open boundary condition (OBC) at half-filling. 
In addition to the aforementioned classical proposals, we also extend the ExcitationSD proposal by incorporating a global spin flip operation, denoted by ExcitationSD+flip. In this proposal, with equal probability, either a random single/double excitation or a global spin flip is performed.

\begin{figure}[htbp]
    \centering
    \includegraphics[width=1.0\textwidth]{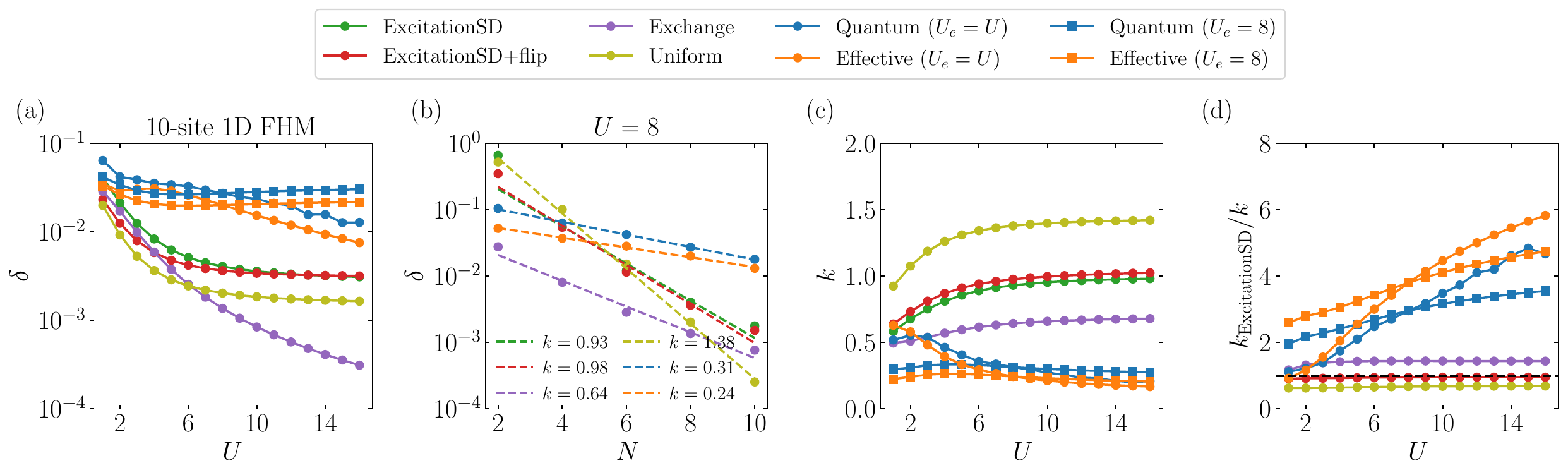}
    \caption{
    The absolute spectral gap $\delta$ obtained by diagonalizing the transition matrix $\mathcal{P}$ 
    of each proposal for the ground state of the 1D FHM. 
    For the Quantum proposal, $\delta$ is obtained as the maximal absolute spectral gap by scanning $\tau$ from $0.1$ to $20$ with a step size of $0.2$. 
    (a) Illustration for $\delta$ of different proposals as a function of $U$ for the $10$-site 1D FHM. (b) $\delta$ of different proposals as a function of the system size $N$ for $U=8$. 
    The function $\delta\approx a 2^{-kN}$ is used to fit the data of each proposal, and the dashed lines are the fitted curves with the obtained $k$ shown in the inset. (c) The fitted exponent $k$ as a function of $U$. (d) $k_{\mathrm{rel}}=k_{\mathrm{ExcitationSD}}/k$ as a function of the parameter $U$. The black dashed line represents $k_{\mathrm{rel}}=1.0$.
    }
    \label{fig: hubbard_model_gap_1D}
\end{figure}

We first analyze the asymptotic behavior for the absolute spectral gaps with the system size $N$ and the on-site interaction 
$U$ for the exact ground state of the one-dimensional (1D) FHM.
For the Quantum proposal, $\delta$ is a function of the evolution time $\tau$.
As shown in Supporting Information, as $\tau$ increases, $\delta$ first 
reaches that of the Effective proposal, denoted by $\delta_{\mathrm{eff}}$, and then oscillates around it.
To examine the best performance that the Quantum proposal can achieve, we
take the maximal absolute spectral gap by scanning $\tau$ from $0.1$ to $20$ with a step size of $0.2$
for each $U$ and $N$. The results obtained with different proposals
are summarized in Figure \ref{fig: hubbard_model_gap_1D},
where we also plot the results obtained by the Quantum proposal with a fixed $U_e=8$ for all $U$.
Figure \ref{fig: hubbard_model_gap_1D}(a) indicates that the Quantum ($U_e=U$) proposal
and that with a fixed $U_e=8$ generally exhibit 
larger spectral gaps $\delta$ than classical proposals for $U \in [1,16]$,
and behave similarly to the corresponding Effective proposals.
Notably, around $U=8$, $\delta$ of the Quantum ($U_e=U$) proposal is approximately an order of magnitude larger than that of the ExcitationSD proposal in the 10-site 1D FHM. 
However, as $U$ increases to infinity, while the absolute spectral gaps of the ExcitationSD, ExcitationSD+flip, and Uniform proposals approach a fixed value, those of the Quantum ($U_e=U$), Effective ($U_e=U$), and Exchange proposals decrease. This is because in the $U=\infty$ limit, Markov chains generated by these proposals become reducible. Using a fixed $U_e=8$ in
the Quantum proposal can avoid this problem, leading to a steady $\delta$ over a wider range of $U$.

Figure \ref{fig: hubbard_model_gap_1D}(b) demonstrates that $\delta$ 
for a fixed value of $U$ exhibits an exponential decay with increasing system size $N$ for all proposals. Following the approach outlined in Ref. \cite{layden_quantum-enhanced_2023},
we fit the data using $\delta(N)=a 2^{-k N}$. Note that both
the prefactor $a$ and the exponent $k$ depend on $U$. 
The Quantum ($U_e=U$) and Effective ($U_e=U$) proposals are found to have the smallest exponents at $U=8$. 
Figure \ref{fig: hubbard_model_gap_1D}(c) presents the obtained exponents 
$k$ for different $U$ using the same fitting procedure, and
Figure \ref{fig: hubbard_model_gap_1D}(d) illustrates the relative 
performance by plotting the ratio $k_{\mathrm{ExcitationSD}}/k$.
We find that for small $U$ ($\approx 1$), the Quantum ($U_e=U$) proposal does not provide advantage over classical proposals. However, it does exhibit an advantage for larger $U$, indicating
the potential for quantum speedup. In comparison, the Quantum approach with a fixed $U_e=8$ shows 
a more balanced performance across all $U$ values. As shown in Supporting Information, 
the advantage of the Quantum proposal in the exponent over classical proposals
persist for 2D and random FHMs.

\begin{figure}[t]
    \centering
    \includegraphics[width=1.0\textwidth]{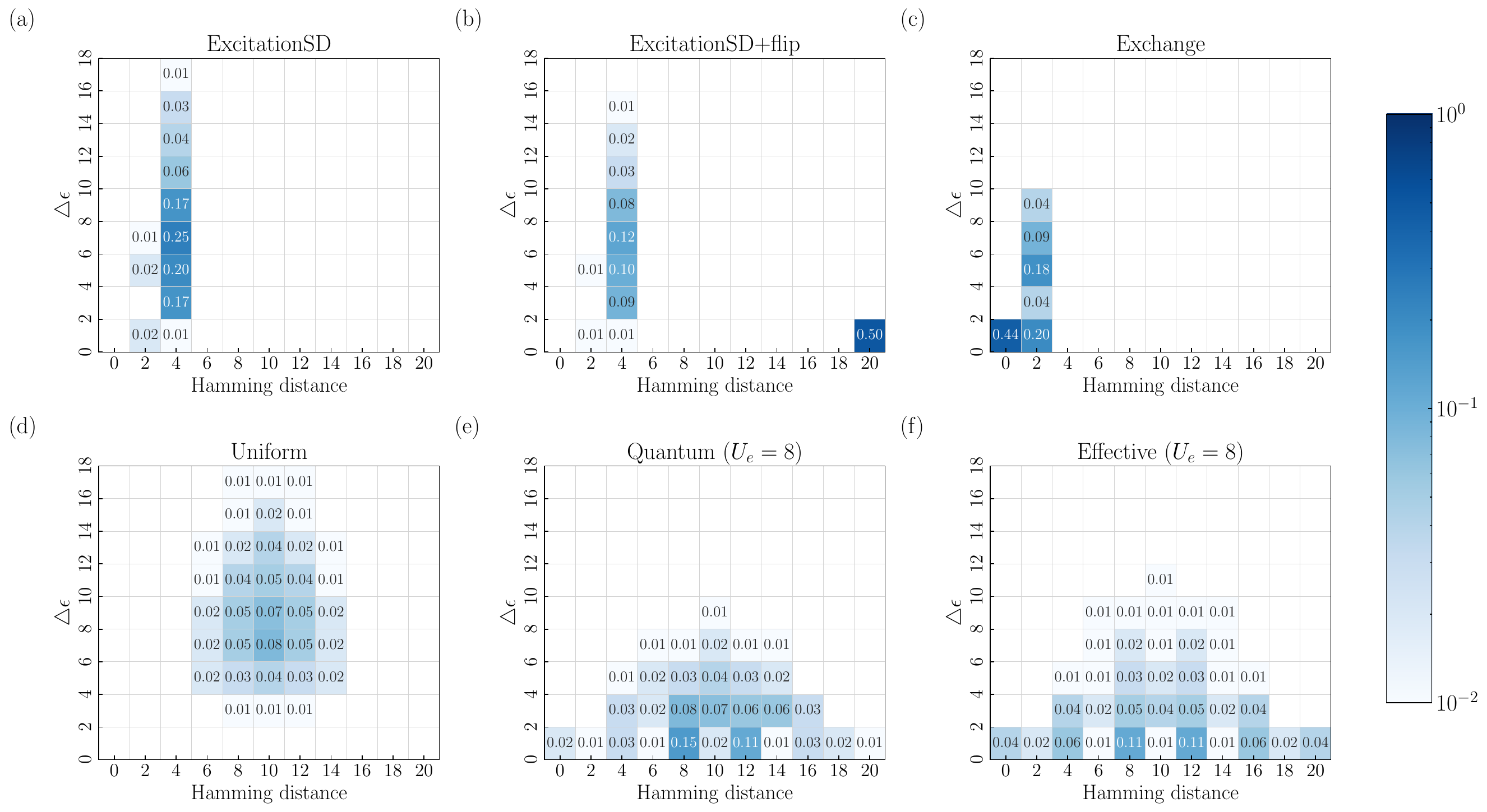}
    \caption{
    Comparison of different proposal probabilities
    $\mathcal{Q}(\boldsymbol{S_i},\cdot)$ from the qubit configuration
    $\boldsymbol{S}_i = (-1,1,1,-1,\cdots, -1,1,1,-1)$ with the largest ground-state probability in 
    the $10$-site 1D FHM with $U=8$. (a)-(f) Two-dimensional histogram of $\mathcal{Q}(\boldsymbol{S_i},\cdot)$ with the Hamming distance
    (between $\boldsymbol{S}_j$ and $\boldsymbol{S}_i$) and the 'energy' gap ($\triangle \epsilon=\mathrm{log}_{10}(P(\boldsymbol{S}_i)/P(\boldsymbol{S}_j))$ as the $x$ and $y$ axes, respectively. (a) ExcitationSD. (b) ExcitationSD+flip. (c) Exchange. (d) Uniform. (e) Quantum ($U_e=8$). (f) Effective ($U_e=8$).}
    \label{fig: hubbard_model_distri_1D}
\end{figure}

To understand how the Quantum proposal speeds up the convergence of the MCMC sampling at larger $U$, 
we introduce the configuration 'energy' defined by
\begin{equation}
    \epsilon(\boldsymbol{S}) = -\log_{10} P(\boldsymbol{S}),\quad
    P(\boldsymbol{S})=|\langle \boldsymbol{S}|\Psi_0\rangle|^2,
\end{equation}
which is analogous to the energy function in the classical Boltzmann distribution.
Specifically, a configuration with high energy $\epsilon(\boldsymbol{S})$ corresponds to a low probability $P(\boldsymbol{S})$, and a large increase in energy 
\begin{eqnarray}
\Delta \epsilon = \epsilon(\boldsymbol{S}_j) - \epsilon(\boldsymbol{S}_i)= \log_{10}(P(\boldsymbol{S}_i)/P(\boldsymbol{S}_j))
\end{eqnarray}
will lead 
to a low acceptance rate in MCMC sampling. 
In Figure 3, we plot the two-dimensional histogram of 
different proposal probabilities $\mathcal{Q}(\boldsymbol{S_i},\cdot)$ 
for the $10$-site 1D FHM with $U=8$,
with the Hamming distance and 'energy' change $\triangle \epsilon$ as
the $x$ and $y$ axes, respectively. 
Here, the qubit configuration $\boldsymbol{S}_i = (-1,1,1,-1,\cdots, -1,1,1,-1)$ is one of the two configurations with the largest ground-state probability (see Supporting Information).
Its spin-flipped counterpart $(1,-1,-1,1,\dots,1,-1,-1,1)$ has an identical probability
due to spin-flip symmetry ($[\hat{H},\hat{U}_{\mathrm{SF}}]=0$, where $\hat{U}_{\mathrm{SF}}=e^{\mathrm{i}\pi(\hat{S}_x-\hat{N}/2)}$), but the largest Hamming distance ($=20$) from $\boldsymbol{S}_i$.
As shown in Figures \ref{fig: hubbard_model_distri_1D}(a)-(c), the ExcitationSD, ExcitationSD+flip,
and Exchange proposals generate configurations that move only by specific Hamming distances. Moreover, the newly generated configurations often exhibit a significant increase in 'energy', leading to a reduced acceptance rate in MCMC sampling. Figure \ref{fig: hubbard_model_distri_1D}(d) shows that although the Uniform proposal allows transitions over unrestricted Hamming distances, it predominantly generates high-energy configurations, thereby also decreasing the MCMC acceptance rate.
In contrast, Figures \ref{fig: hubbard_model_distri_1D}(e) and (f) demonstrate that the Quantum and Effective proposals can generate configurations with a range of Hamming distances while maintaining relatively low 'energy'. This distinctive property significantly enhances Markov chain convergence, differentiating quantum moves from classical moves.

\begin{figure}[t]
    \centering
    \includegraphics[width=1.0\textwidth]{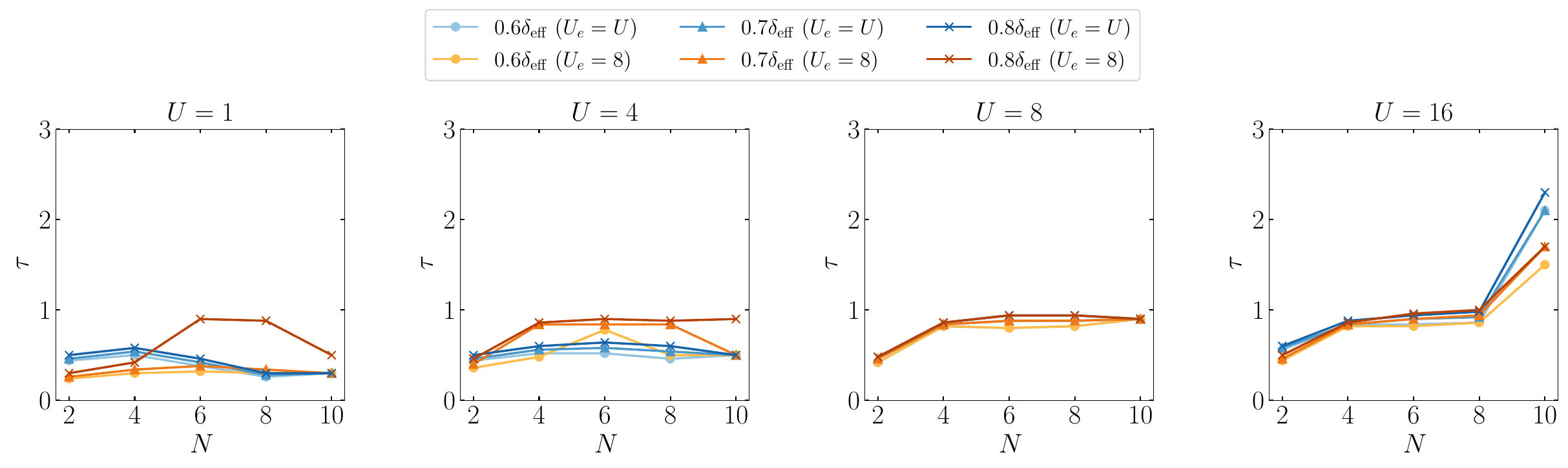}
    \caption{The evolution time $\tau$ required for $\delta$ of the Quantum proposal to first
    exceeds $c\delta^{\mathrm{eff}}$ ($c=0.6$, 0.7, and 0.8) as a function of the system size $N$ for the ground state of 1D FHMs with different $U$.
    }
    \label{fig: hubbard_model_time_need_1D}
\end{figure}

As discussed in the previous section, it is also crucial to examine the asymptotic behavior of the runtime $t_{s,q}$ in order to assess whether the Quantum proposal can achieve a quantum advantage in computational time.
The runtime $t_{s,q}$ of a single quantum move is proportional to the evolution time $\tau$.
Here, we consider the evolution time required to first reach a certain fraction of $\delta_\mathrm{eff}$ and analyze its dependence on the system size. This is motivated by the observation that as the evolution time increases, the spectral gap of the Quantum proposal oscillates around $\delta_\mathrm{eff}$ (see
Supporting Information for details).
Figure \ref{fig: hubbard_model_time_need_1D} shows the evolution time $\tau$ at which $\delta$
of the Quantum proposals ($U_e=U$ and $U_e=8$) first exceeds $c\delta_\mathrm{eff}$ for $c=0.6$, 0.7, and 0.8, respectively. Notably, the required evolution time does not increase rapidly with system size.
In particular, it reaches a plateau for both $U=4$ and $U=8$.
Similar behaviors are also observed for 2D FHMs shown in Supporting Information.
Based on Eq. \eqref{eq:classicalVsQuantum}, these findings suggest that the Quantum proposal, with an appropriately chosen parameter $U_e$, may offer a potential quantum speedup over classical proposals for sufficiently large systems.

\begin{figure}[t]
    \centering
    \includegraphics[width=1.0\textwidth]{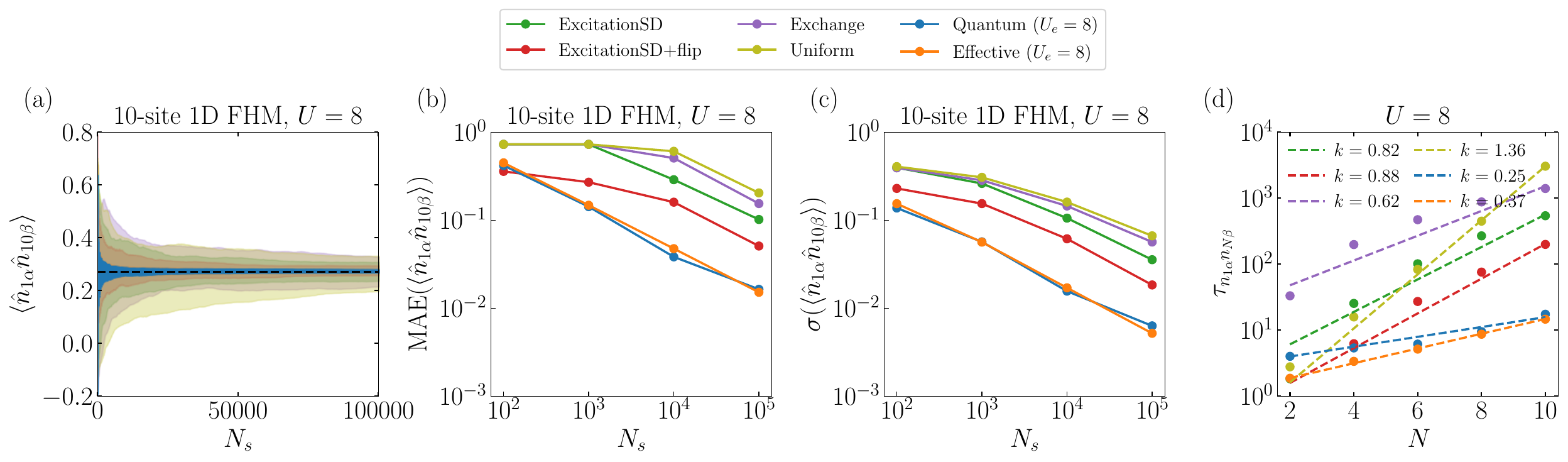}
    \caption{
    Estimation of an observable $\langle\hat{n}_{1\alpha}\hat{n}_{N\beta}\rangle$ by 100 independent Markov chains with different proposals for the exact ground state of 1D FHM with $U=8$. (a) The distribution of the estimated $\langle\hat{n}_{1\alpha}\hat{n}_{10\beta}\rangle$ for a given $N_s$ with different proposals. The black dashed line represents the exact value.
    (b) maximum absolute errors (MAE) for the estimated $\langle\hat{n}_{1\alpha}\hat{n}_{10\beta}\rangle$ as a function of $N_s$. (c) standard deviation $\sigma$ for the estimated $\langle\hat{n}_{1\alpha}\hat{n}_{10\beta}\rangle$ as a function of $N_s$. (d) Estimated $\tau_{n_{1\alpha} n_{10\beta}}$ as a function of $N$ for different proposals using the MCMC algorithm with $N_s = 10^7$. The data were further fitted using $a2^{kN}$ (dashed lines) with the obtained
    exponents shown in the inset.
    }
    \label{fig: hubbard_model_1D_MCMC}
\end{figure}

To further assess the quality of samples generated by different proposals, we evaluate an observable $\langle\hat{n}_{1\alpha}\hat{n}_{N\beta}\rangle$ using the MCMC algorithm for the exact ground state of
the 10-site 1D FHM with $U = 8$. Figure \ref{fig: hubbard_model_1D_MCMC}(a) presents the results of 100 independent Markov chains for each proposal. The Quantum proposal demonstrates superior performance, yielding more accurate results with smaller variations for a given sample size $N_s$. Compared to the best classical proposal (ExcitationSD+flip) for this observable,
the Quantum proposal reduces the maximum error and standard deviation by approximately a factor of 3 for $N_s=10^5$, 
as shown in Figures \ref{fig: hubbard_model_1D_MCMC}(b) and (c). 
This improvement suggests that the effective sample size $N_{\mathrm{eff}}$ 
is roughly 9 times larger, which aligns well with the estimated integrated autocorrelation time 
$\tau_{n_{1\alpha}n_{N\beta}}$ for $N=10$ depicted in Figure \ref{fig: hubbard_model_1D_MCMC}(d). 
We extend the same analysis to other system sizes and
fit the obtained $\tau_{n_{1\alpha}n_{N\beta}}$ as a function of $N$ 
using $a 2^{kN}$ in Figure \ref{fig: hubbard_model_1D_MCMC}(d). 
The results reveal that the Quantum proposal exhibits the smallest $k$,
and hence the slowest increase in 
$\tau_{n_{1\alpha}n_{N\beta}}$ as the system size $N$ increases, 
which is consistent with the trend observed for the absolute spectral gap. 
This further underscores the higher quality of samples produced by the Quantum proposal.

\begin{figure}[t]
    \centering
    \includegraphics[width=0.6\textwidth]{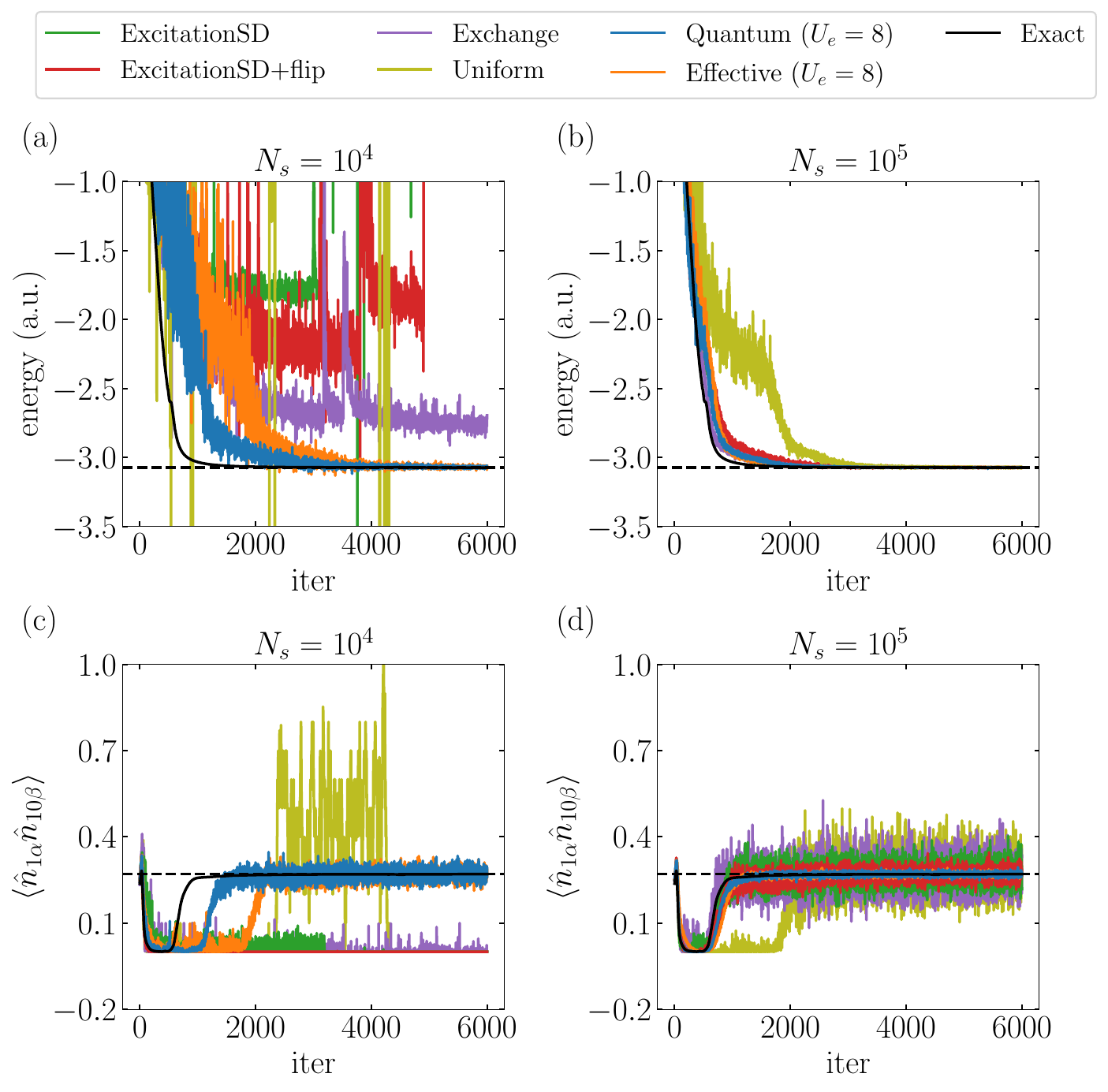}
    \caption{
    The VMC optimization process of different proposals using the RBM ($\alpha=3$) ansatz for 10-site 1D FHM with $U=8$. (a),(b) energy, (c),(d) $\langle\hat{n}_{1\alpha}\hat{n}_{10\beta}\rangle$. 
    Black solid lines in (a) and (b) represent the optimization trajectory using the exact gradients without sampling. Black dashed lines represent the exact ground-state energy in (a) and (b) or $\langle\hat{n}_{1\alpha}\hat{n}_{10\beta}\rangle$ for the exact ground state in (c) and (d).}
    \label{fig: hubbard_model_VMC_1D}
\end{figure}

Finally, we illustrate the performance of the QA-VMC algorithm in practical applications 
by combining it with the RBM ansatz ($\alpha=3$) to target the ground-state of the
10-site 1D FHM with $U=8$. The results obtained using two different sample sizes ($N_s=10^4$ and $N_s=10^5$)
are presented in Figure \ref{fig: hubbard_model_VMC_1D}. 
Figure \ref{fig: hubbard_model_VMC_1D}(a) and (b) demonstrate that the variational energy computed by QA-VMC 
converges more efficiently toward the exact ground-state energy, requiring fewer samples $N_s$ compared with classical proposals. Specifically, VMC with classical proposals fail to converge to the correct
ground state using $N_s=10^4$. In contrast, the convergence trajectory of QA-VMC aligns more closely with
the optimization using the exact gradients (black lines), 
highlighting its superior efficiency due to the higher quality of samples.
Additionally, Figure \ref{fig: hubbard_model_VMC_1D}(c) and (d) display the estimated $\langle\hat{n}_{1\alpha}\hat{n}_{10\beta}\rangle$ during the VMC optimizations. 
The results obtained with the Quantum proposals are found to
exhibit better accuracy and 
smaller oscillations at the same sample size $N_s$ 
compared with classical proposals.
This shows the potential of QA-VMC for significantly enhancing the 
performance of the VMC algorithm for large systems.

\subsection{Molecules}
After benchmarking QA-VMC for FHMs across various system sizes and interaction parameters,
we now apply it to chemical systems with more realistic interactions.
The molecular Hamiltonian is given by
\begin{eqnarray}
    \hat{H} = \sum_{\substack{pq,\sigma}} h_{pq} \hat{a}_{p\sigma}^\dagger \hat{a}_{q\sigma} 
+ \frac{1}{2} \sum_{\substack{pqrs,\sigma\tau}} g_{pqrs} \hat{a}_{p\sigma}^\dagger \hat{a}_{r\tau}^\dagger \hat{a}_{s\tau} \hat{a}_{q\sigma} 
+ E_{\mathrm{nuc}},
\end{eqnarray}
where $h_{pq}$ and $g_{pqrs}$ denote the one- and two-electron molecular integrals, respectively, and $E_{\mathrm{nuc}}$ represents the nuclear repulsion energy. The Fermionic Hamiltonian is then 
transformed into a qubit Hamiltonian via the Jordan-Wigner mapping\cite{jordan1928pauli}
for subsequent studies.
Analogous to the effective Hamiltonian approach employed in the Hubbard model, we can construct an effective Hamiltonian for molecular systems by fixing the bond length $R_e$, and the resulting proposal will
be defined by the Quantum ($R_e$) proposal.

Additionally, we introduce another way
to design an effective Hamiltonian, by incorporating an artificial hopping term into the Hamiltonian, viz.,
\begin{eqnarray}
    \hat{H}(\gamma_e) = \big((1-\gamma_e)\sum_{\substack{pq,\sigma}} h_{pq} \hat{a}_{p\sigma}^\dagger \hat{a}_{q\sigma} +\gamma_e\alpha\hat{H}_{\mathrm{hopping}}\big)
+ \frac{1}{2} \sum_{\substack{pqrs,\sigma\tau}} g_{pqrs} \hat{a}_{p\sigma}^\dagger \hat{a}_{r\tau}^\dagger \hat{a}_{s\tau} \hat{a}_{q\sigma} 
+ E_{\mathrm{nuc}},\label{eq: ham_hopping}
\end{eqnarray}
where $\hat{H}_{\mathrm{hopping}}$ is
\begin{eqnarray}
\hat{H}_{\mathrm{hopping}}=-\sum_{p\ne q, \sigma}\hat{a}_{p\sigma}^\dagger \hat{a}_{q\sigma}.
\end{eqnarray}
Here, $\gamma_e\in[0.0, 1.0]$ is a tunable parameter that governs the relative contribution of the hopping term in the one-body part,
and the normalization factor $\alpha=\frac{\lVert h\lVert_{\mathrm{F}}}{\sqrt{n(n-1)}}$, where $n$ is the number of spatial orbitals, 
ensures appropriate scaling of the one-electron component. 
We will denote the Quantum proposal using Eq. \eqref{eq: ham_hopping}
by the Quantum (hopping, $\gamma_e$) proposal.
When $\gamma_e=0.0$, the Quantum (hopping, $\gamma_e=0.0$) proposal reduces to the Quantum ($R_e=R$) proposal. Conversely, setting $\gamma_e=1.0$ replaces the entire one-body term with the hopping operator. As the optimal value of $\gamma_e$ is generally unknown, we adopt a stochastic strategy as in Ref. \cite{layden_quantum-enhanced_2023}, in which $\gamma_e$ is sampled from a uniform distribution in the interval $[0.1,0.4]$. The resulting proposal
will be denoted by the Quantum (hopping, random) proposal (see Supporting
Information for details of implementation).

\subsubsection{Hydrogen chains}
\begin{figure}[t]
    \centering
    \includegraphics[width=1.0\textwidth]{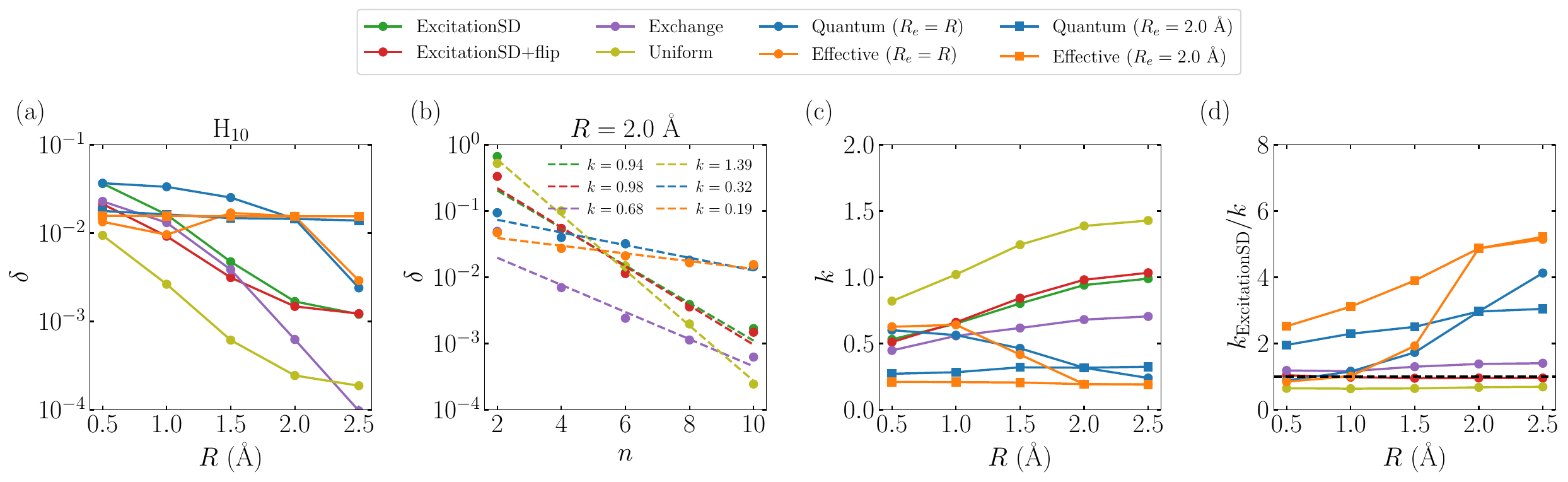}
    \caption{
    The absolute spectral gap $\delta$ obtained by diagonalizing the transition matrix $\mathcal{P}$ 
    of each proposal for the ground state of the hydrogen chains H$_n$. 
    For the Quantum proposal, $\delta$ is obtained as the maximal absolute spectral gap by scanning $\tau$ from $0.1$ to $60.0$ with a step size of $0.2$. 
    (a) $\delta$ of different proposals as a function of $R$ for \ce{H10}. (b) $\delta$ of different proposals as a function of the system size $n$ at $R=2.0$ \AA.
    The function $a 2^{-kN}$ is used to fit the data of each proposal, and the dashed lines are the fitted curves with the obtained $k$ shown in the inset. (c) The fitted exponent $k$ as a function of $R$. (d) $k_{\mathrm{rel}}=k_{\mathrm{ExcitationSD}}/k$ as a function of the parameter $U$. The black dashed line represents $k_{\mathrm{rel}}=1.0$.
    }
    \label{fig: Hn_gap_all}
\end{figure}

A typical example, closely related to FHMs, is the hydrogen chains at varying interatomic distances $R$, which can undergo transitions from weakly correlated systems at small $R$ to strongly correlated systems at
larger $R$. We employed the orthonormalized atomic orbitals (OAO) 
obtained with the STO-3G basis. 
Figure \ref{fig: Hn_gap_all} presents the absolute spectral gaps $\delta$ obtained 
with different proposals for the ground state of hydrogen chains $\ce{H}_n$.  
As depicted in Figure \ref{fig: Hn_gap_all}(a), as the bond length $R$ increases from $0.5~\text{\AA}$ to $2.5~\text{\AA}$, 
the absolute spectral gap $\delta$ for the Quantum ($R_e = R$) proposal is generally much greater 
than those of classical proposals. Similar to FHMs in the large $U$ limit, $\delta$ for the Exchange, Quantum $(R_e=R)$, and Effective $(R_e=R)$ proposals decreases to zero as $R$ increases, due to the lost of
irreducibility for the generated Markov chains in the $R=\infty$
limit. In contrast, other proposals maintain a nonzero $\delta$ at large $R$. 
In particular, by fixing $R_e$ to a specific value, 
such as 2.0~\AA, the spectral gap of the Quantum proposal
can sustain a large value across different $R$, see Figure \ref{fig: Hn_gap_all}(a).
Figure \ref{fig: Hn_gap_all}(b) shows that $\delta$ decays exponentially with system size and is well-fitted by the function $a 2^{-kN}$. At $R=2.0~\text{\AA}$, the fitted exponent $k$ for the Quantum proposal is only about one-third of that of the widely used ExcitationSD proposal, indicating a significant potential speedup for large systems.
Figure \ref{fig: Hn_gap_all}(c) and (d) display the fitted exponents $k$ for different bond lengths and the relative exponents
$k_{\mathrm{rel}} = k_{\mathrm{ExcitationSD}}/k$ 
compared against that of ExcitationSD, respectively. It is evident that at larger $R>1.5~\text{\AA}$, 
where the ground-state configurations become more concentrated on some configurations
separated by large Hamming distances (see Supporting Information), 
the Quantum proposals start to outperform classical proposals. 

\begin{figure}[t]
    \centering
    \includegraphics[width=1.0\textwidth]{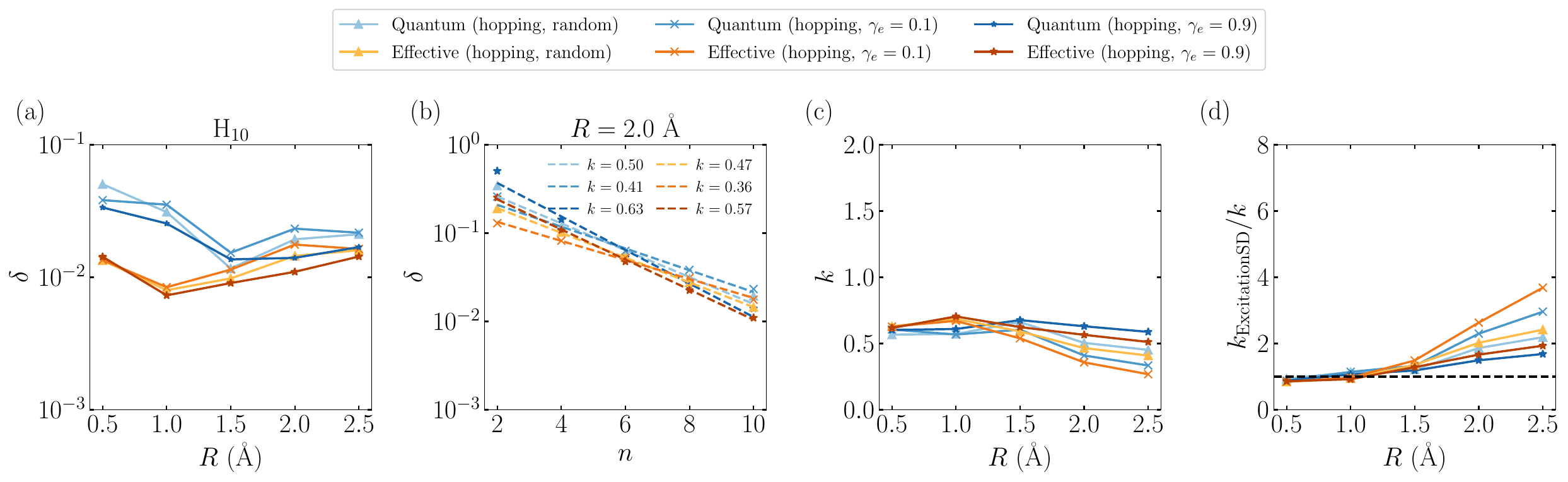}
    \caption{
    The absolute spectral gap $\delta$ obtained by diagonalizing the transition matrix $\mathcal{P}$ 
    of different Quantum (hopping) proposals for the ground state of the hydrogen chains H$_n$. 
    For different Quantum (hopping) proposals, $\delta$ is obtained as the maximal absolute spectral gap by scanning $\tau$ from $0.1$ to $60.0$ with a step size of $0.2$. 
    (a) $\delta$ of different Quantum (hopping) proposals as a function of $R$ for \ce{H10}. (b) $\delta$ of different Quantum (hopping) proposals as a function of the system size $n$ at $R=2.0~\text{\AA}$.
    The function $a 2^{-kN}$ is used to fit the data of each proposal, and the dashed lines are the fitted curves with the obtained $k$ shown in the inset. (c) The fitted exponent $k$ as a function of $R$. (d) $k_{\mathrm{rel}}=k_{\mathrm{ExcitationSD}}/k$ as a function of the parameter $U$. The black dashed line represents $k_{\mathrm{rel}}=1.0$.
    }
    \label{fig: Hn_delta_gap_gamma}
\end{figure}

\begin{figure}[t]
    \centering
    \includegraphics[width=0.6\textwidth]{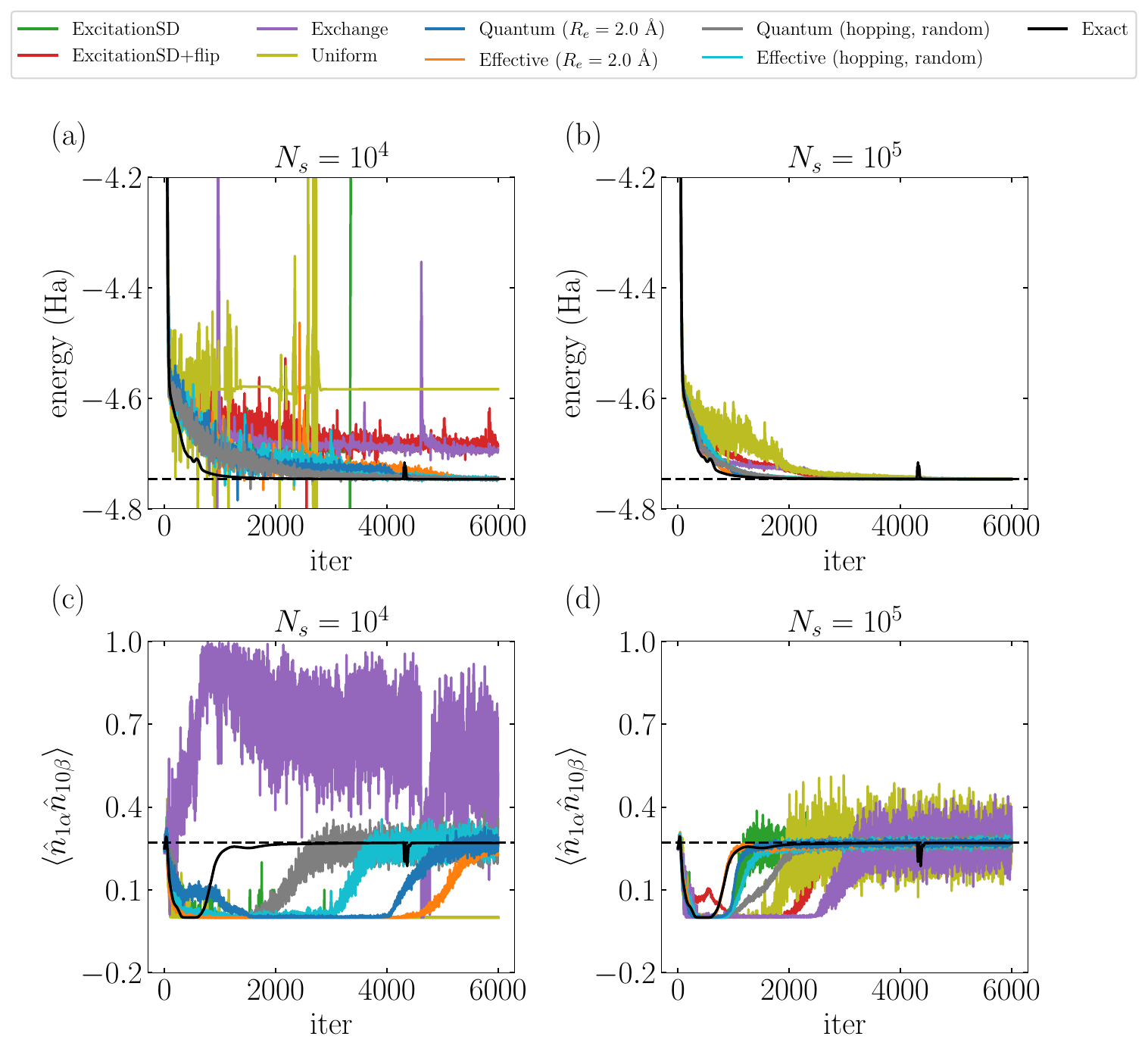}
    \caption{
    The VMC optimization process of different proposals using the RBM ($\alpha=3$) ansatz for the hydrogen
    chain \ce{H10} with $R=2.0$ \AA. (a),(b) energy, (c),(d) $\langle\hat{n}_{1\alpha}\hat{n}_{10\beta}\rangle$. 
    Black solid lines in (a) and (b) represent the optimization trajectory using the exact gradients without sampling. Black dashed lines represent the exact ground-state energy in (a) and (b) or $\langle\hat{n}_{1\alpha}\hat{n}_{10\beta}\rangle$ for the exact ground state in (c) and (d).
    }
    \label{fig: Hn_VMC}
\end{figure}


Figure \ref{fig: Hn_delta_gap_gamma} presents the absolute spectral gaps $\delta$ obtained from both the Quantum and Effective proposals, using the effective Hamiltonian $\hat{H}(\gamma_e)$ defined in Eq. \eqref{eq: ham_hopping}.
In Figure \ref{fig: Hn_delta_gap_gamma}(a), the Quantum (hopping) proposals exhibit robust performance with the bond length $R$ from $0.5~\text{\AA}$ to $2.5~\text{\AA}$. In particular, in contrast to the quantum ($R_e=R$) proposal, the absolute spectral gap $\delta$ for the Quantum (hopping) proposals remains nonvanishing at large bond lengths, indicating improved performance in the dissociation limit.
Figure \ref{fig: Hn_delta_gap_gamma}(b) shows that a small value of $\gamma_e$ results in a favorable scaling factor $k$ at $R=2.0~\text{\AA}$. Additionally, the Quantum (hopping, random) proposal achieves a similarly small $k$, highlighting the effectiveness of the randomized strategy.
A more detailed comparison of the scaling behavior is shown in Figure \ref{fig: Hn_delta_gap_gamma}(c), where a clear hierarchy emerges: $k~(\gamma_e=0.1) < k~(\mathrm{random}) < k~(\gamma_e=0.9)$.
Finally, Figure \ref{fig: Hn_delta_gap_gamma}(d) demonstrates that the Quantum (hopping, random) proposal consistently outperforms classical approaches for bond lengths greater than $1.5~\text{\AA}$, underscoring its advantage in strongly correlated regimes.

Finally, we illustrate the performance of QA-VMC combined with the RBM ansatz ($\alpha=3$) 
for computing the ground state of the hydrogen chain $\ce{H_{10}}$ and the observable $\langle\hat{n}_{1\alpha}\hat{n}_{10\beta}\rangle$ at $R=2.0~\text{\AA}$. 
The estimated energy and $\langle\hat{n}_{1\alpha}\hat{n}_{10\beta}\rangle$ during the 
optimization process are shown in Figure \ref{fig: Hn_VMC} for two different sample sizes,
$N_s=10^4$ and $N_s=10^5$. For small $N_s$, Figure \ref{fig: Hn_VMC}(a) and (c) reveal 
that the Quantum ($R_e=2.0~\text{\AA}$) and Quantum (hopping) proposals significantly outperform
classical proposals. 
Similar to the case for FHMs, VMC with classical proposals all fail to
converge to the correct ground state and $\langle\hat{n}_{1\alpha}\hat{n}_{10\beta}\rangle$
for $N_s=10^4$. Only when $N_s$ is increased to $10^5$, classical proposals begin to converge to the correct results. 
These results are consistent with the findings for FHMs, and
underscore the potential of QA-VMC to accelerate VMC for molecular systems.

\subsubsection{Water molecule}
\begin{figure}[t]
    \centering
    \includegraphics[width=0.6\textwidth]{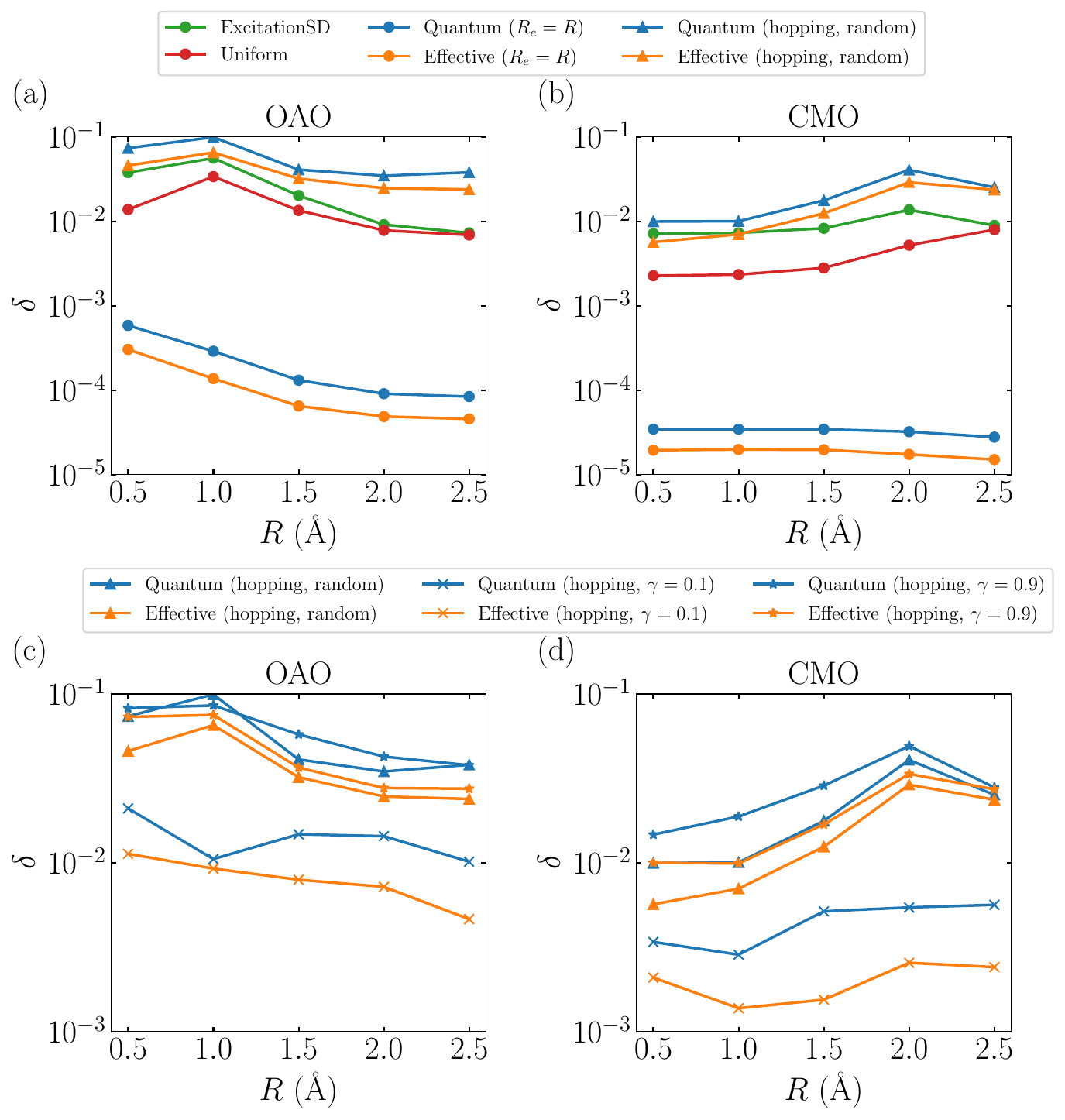}
    \caption{The absolute spectral gap $\delta$ obtained by diagonalizing the transition matrix $\mathcal{P}$ 
    of each proposal for the ground state of \ce{H2O}. 
    For the Quantum proposal, $\delta$ is obtained as the maximal absolute spectral gap by scanning $\tau$ from $0.1$ to $40.0$ with a step size of $0.2$. $\delta$ for the classical proposals, Quantum and Effective proposals with or without the hopping term in the OAO (a) or CMO (b) basis. $\delta$ for the Quantum (hopping) and the Effective (hopping) proposals with different $\gamma_e$ selections in the OAO (c) or CMO (d).}
    \label{fig: H2O_gap_all}
\end{figure}

In addition to hydrogen chains, we also consider the water molecule
as an example for more realistic systems.
Here, we fix the bond angle at $\angle \ce{H-O-H} = 104.5^\circ$ and vary the \ce{O-H} bond length $R$. As $R$ increases, it also exhibits a
transition from a weakly correlated to a strongly correlated system.
Figures \ref{fig: H2O_gap_all}(a) and (b) show the absolute spectral gap $\delta$ obtained using various proposals for the ground state of \ce{H2O} at different $R$ using OAOs and canonical molecular orbitals (CMOs). The results indicate that the Quantum and Effective proposals incorporating hopping terms outperform the classical proposals, whereas the classical proposals exhibit better performance than the Quantum and Effective proposals without hopping.
This behavior can be attributed to the heterogeneous atomic composition of the system, because in such case some high-energy excited states are dominated by a very few configurations. Consequently, as shown by
Eq. \eqref{eq: Effective_proposal}, if a configuration is transition to 
an excited state dominated by a very few configurations, then it is hard to transit to other configurations by the Quantum proposals and the mixing time is therefore increased. The detailed mechanism is discussed in the Supplementary Information. Including the hopping term in the effective Hamiltonian can increase the transition probability, thereby substantially enhancing the absolute spectral gap $\delta$.
Figures \ref{fig: H2O_gap_all}(c) and (d) further reveal that the absolute spectral gap $\delta$ obtained from the Quantum (hopping, $\gamma_e = 0.9$) proposal is significantly larger than that from the Quantum (hopping, $\gamma_E = 0.1$) proposal. This suggests that for \ce{H2O}, increasing the proportion of hopping terms in the one-body part of the effective Hamiltonian is beneficial for enhancing the absolute spectral gap. In addition, the Quantum (hopping, random) proposal also achieves comparably favorable performance.

\begin{figure}[t]
    \centering
    \includegraphics[width=0.6\textwidth]{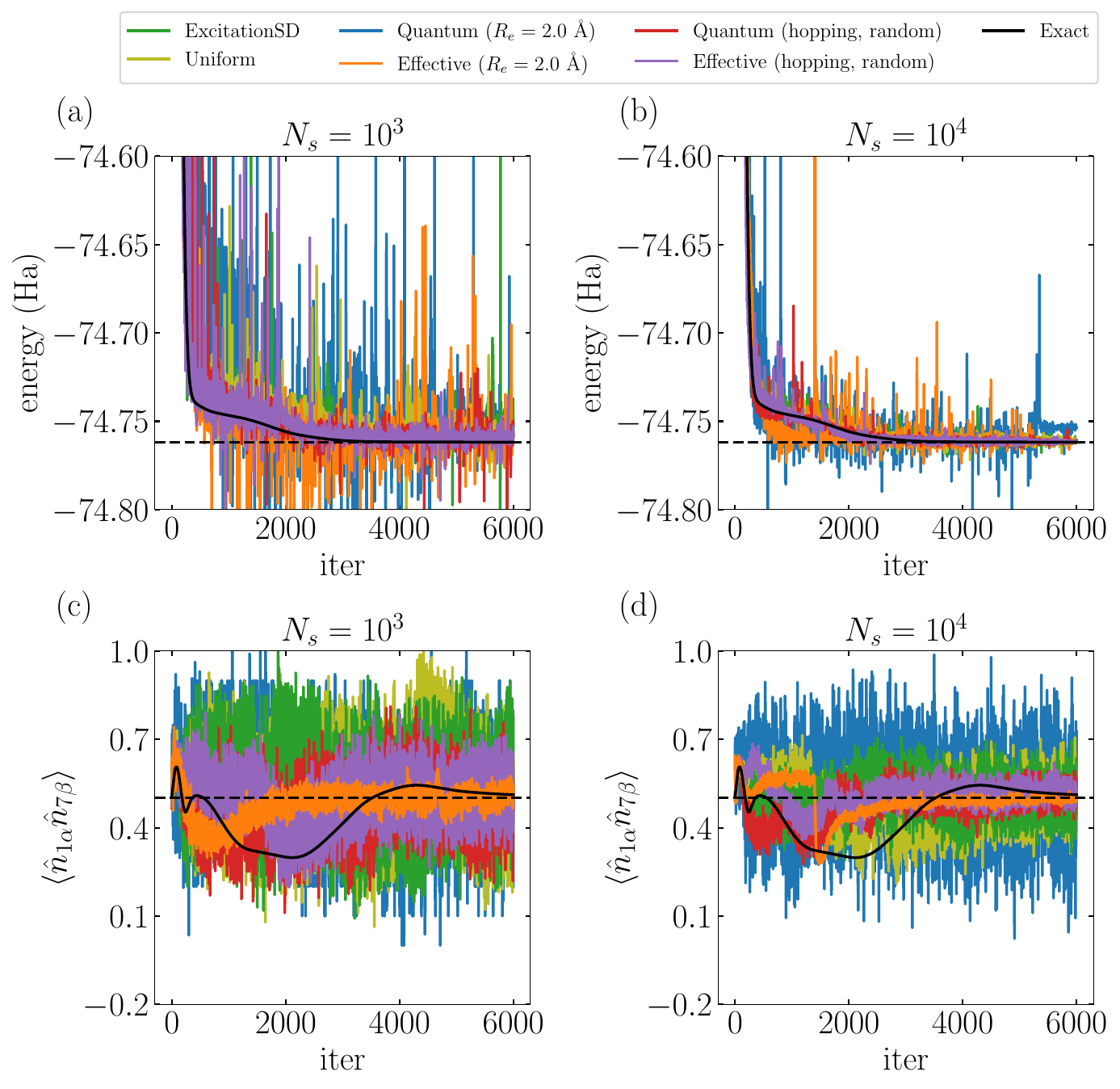}
    \caption{The VMC optimization process of different proposals using the RBM ($\alpha=3$) ansatz for \ce{H2O} at a bond length $R=2.0$ \AA. (a),(b) energy, (c),(d) $\langle\hat{n}_{1\alpha}\hat{n}_{7\beta}\rangle$. 
    Black solid lines in (a) and (b) represent the optimization trajectory without sampling. Black dashed lines represent the exact ground-state energy in (a) and (b) or $\langle\hat{n}_{1\alpha}\hat{n}_{7\beta}\rangle$ for the exact ground state in (c) and (d).}
    \label{fig: H2O_VMC}
\end{figure}

Finally, we examine the performance of QA-VMC combined with the RBM ansatz ($\alpha = 3$) in computing both the ground-state energy of \ce{H2O} and an illustrative observable $\langle \hat{n}_{1\alpha} \hat{n}_{7\beta} \rangle$ at a bond length $R = 2.0~\text{\AA}$ and a fixed bond angle $\angle \ce{H-O-H} = 104.5^\circ$. The estimated energy and the expectation value $\langle \hat{n}_{1\alpha} \hat{n}_{7\beta} \rangle$ during the optimization process are shown in Figure \ref{fig: H2O_VMC} for two different sample sizes ($N_s = 10^3$ and $N_s = 10^4$). 
Figures \ref{fig: H2O_VMC}(a) and (b) show that the variational energy curves obtained from QA-VMC and conventional VMC using classical proposals exhibit no significant differences. In contrast, Figures \ref{fig: H2O_VMC}(c) and (d) reveal that the estimation of $\langle \hat{n}_{1\alpha} \hat{n}_{7\beta} \rangle$ using the Quantum and Effective proposals with hopping terms is significantly more accurate than that obtained with classical proposals, featuring smaller fluctuations. 
Interestingly, the Quantum ($R_e = 2.0~\text{\AA}$) proposal results in strong oscillations, whereas the Effective ($R_e = 2.0~\text{\AA}$) proposal demonstrates the opposite behavior, achieving notably stable and accurate results.
This performance appears to contradict with the small absolute spectral gap $\delta$ observed for the Effective ($R_e = 2.0~\text{\AA}$) proposal in Figure \ref{fig: H2O_gap_all}. However, this apparent contradiction arises because the dominant electronic configuration in the highest excited state of the Hamiltonian, where the five highest-energy orbitals are all doubly occupied, has a low probability under the VMC sampling
when the optimization starts from a reasonable initial configuration. In contrast, the mixing time shown in Eq. 
\eqref{eq: mixing_time_def}, which measures the worst-case 
scenario, is affected by such configuration.

\section{Conclusion}
In this work, inspired by the QeMCMC algorithm\cite{layden_quantum-enhanced_2023}, originally designed for sampling classical Boltzmann distributions of spin models, we introduced the QA-VMC algorithm for solving the ground state of quantum many-body problems by leveraging the capabilities of quantum computers to enhance the sampling efficiency in VMC simulations. 
Pilot applications to FHMs and molecular systems reveal that the Quantum proposal exhibits larger absolute spectral gaps and reduced autocorrelation times compared to classical proposals, leading to more efficient sampling and faster convergence to the ground state in VMC. 
This advantage is found to be especially pronounced for specific parameter ranges, where the ground-state configurations are concentrated in some dominant configurations separated by large Hamming distances.
Besides, we also identified limitations of the introduced Quantum proposal, particularly when the system parameters approach some extreme values, leading to reducible Markov chains and vanishing absolute spectral gaps. To mitigate these issues, we proposed fixing certain parameters in the Hamiltonian used for
time evolution in the Quantum proposal, which can maintain a non-zero absolute spectral gap and
exhibit advantages over classical proposals across a wider range of system parameters and sizes.
Furthermore, to extend the applicability of QA-VMC for molecular systems, we propose the Quantum (hopping) proposal, which incorporates additional hopping terms into the molecular Hamiltonian. This approach offers greater generality for molecular systems and is capable of delivering results comparable to those of the Quantum proposal for hydrogen chains.
Our results suggest that QA-VMC has the potential to enhance the performance of VMC algorithms for large systems. 
By providing samples of good quality,
less samples can be used in the VMC optimization, which also
reduces the computational cost for the evaluation of local energy
and gradients.

Future work will focus on further optimizing the Quantum proposal, including the automatic optimization of the evolution time, the use of Trotter decomposition
or other Hamiltonian simulation techniques, and investigating the performance of QA-VMC on noisy quantum simulators and real quantum hardware. Additionally, exploring the application of QA-VMC to other quantum systems with more complex Hamiltonians will be crucial for assessing its broader applicability and potential for quantum advantage.

\begin{acknowledgement}
We acknowledge helpful discussions with Ming Gong. This work was supported by the Innovation Program for Quantum Science and Technology (Grant No. 2023ZD0300200) and the Fundamental Research Funds for the Central Universities.
\end{acknowledgement}

\begin{suppinfo}
Time-averaged Quantum proposal versus Effective proposal, 
the absolute spectral gap of the Quantum proposal as a function of the evolution time,
exact ground state distributions of the investigated models,
and additional results for Fermi-Hubbard models, hydrogen chains, and \ce{H2O}.



\end{suppinfo}

\bibliography{main}

\providecommand{\latin}[1]{#1}
\makeatletter
\providecommand{\doi}
  {\begingroup\let\do\@makeother\dospecials
  \catcode`\{=1 \catcode`\}=2 \doi@aux}
\providecommand{\doi@aux}[1]{\endgroup\texttt{#1}}
\makeatother
\providecommand*\mcitethebibliography{\thebibliography}
\csname @ifundefined\endcsname{endmcitethebibliography}
  {\let\endmcitethebibliography\endthebibliography}{}
\begin{mcitethebibliography}{72}
\providecommand*\natexlab[1]{#1}
\providecommand*\mciteSetBstSublistMode[1]{}
\providecommand*\mciteSetBstMaxWidthForm[2]{}
\providecommand*\mciteBstWouldAddEndPuncttrue
  {\def\EndOfBibitem{\unskip.}}
\providecommand*\mciteBstWouldAddEndPunctfalse
  {\let\EndOfBibitem\relax}
\providecommand*\mciteSetBstMidEndSepPunct[3]{}
\providecommand*\mciteSetBstSublistLabelBeginEnd[3]{}
\providecommand*\EndOfBibitem{}
\mciteSetBstSublistMode{f}
\mciteSetBstMaxWidthForm{subitem}{(\alph{mcitesubitemcount})}
\mciteSetBstSublistLabelBeginEnd
  {\mcitemaxwidthsubitemform\space}
  {\relax}
  {\relax}

\bibitem[Martin \latin{et~al.}(2016)Martin, Reining, and
  Ceperley]{martin_interacting_2016}
Martin,~R.~M.; Reining,~L.; Ceperley,~D.~M. \emph{Interacting {Electrons}:
  {Theory} and {Computational} {Approaches}}; Cambridge University Press:
  Cambridge, 2016\relax
\mciteBstWouldAddEndPuncttrue
\mciteSetBstMidEndSepPunct{\mcitedefaultmidpunct}
{\mcitedefaultendpunct}{\mcitedefaultseppunct}\relax
\EndOfBibitem
\bibitem[Hohenberg and Kohn(1964)Hohenberg, and
  Kohn]{hohenberg_inhomogeneous_1964}
Hohenberg,~P.; Kohn,~W. Inhomogeneous {Electron} {Gas}. \emph{Phys. Rev.}
  \textbf{1964}, \emph{136}, B864--B871\relax
\mciteBstWouldAddEndPuncttrue
\mciteSetBstMidEndSepPunct{\mcitedefaultmidpunct}
{\mcitedefaultendpunct}{\mcitedefaultseppunct}\relax
\EndOfBibitem
\bibitem[Kohn and Sham(1965)Kohn, and Sham]{kohn_self-consistent_1965}
Kohn,~W.; Sham,~L.~J. Self-{Consistent} {Equations} {Including} {Exchange} and
  {Correlation} {Effects}. \emph{Phys. Rev.} \textbf{1965}, \emph{140},
  A1133--A1138\relax
\mciteBstWouldAddEndPuncttrue
\mciteSetBstMidEndSepPunct{\mcitedefaultmidpunct}
{\mcitedefaultendpunct}{\mcitedefaultseppunct}\relax
\EndOfBibitem
\bibitem[Runge and Gross(1984)Runge, and Gross]{runge_density-functional_1984}
Runge,~E.; Gross,~E. K.~U. Density-{Functional} {Theory} for {Time}-{Dependent}
  {Systems}. \emph{Phys. Rev. Lett.} \textbf{1984}, \emph{52}, 997--1000\relax
\mciteBstWouldAddEndPuncttrue
\mciteSetBstMidEndSepPunct{\mcitedefaultmidpunct}
{\mcitedefaultendpunct}{\mcitedefaultseppunct}\relax
\EndOfBibitem
\bibitem[Čížek(1966)]{cizek_correlation_1966}
Čížek,~J. On the {Correlation} {Problem} in {Atomic} and {Molecular}
  {Systems}. {Calculation} of {Wavefunction} {Components} in {Ursell}‐{Type}
  {Expansion} {Using} {Quantum}‐{Field} {Theoretical} {Methods}. \emph{J.
  Chem. Phys.} \textbf{1966}, \emph{45}, 4256--4266\relax
\mciteBstWouldAddEndPuncttrue
\mciteSetBstMidEndSepPunct{\mcitedefaultmidpunct}
{\mcitedefaultendpunct}{\mcitedefaultseppunct}\relax
\EndOfBibitem
\bibitem[Purvis and Bartlett(1982)Purvis, and Bartlett]{purvis_full_1982}
Purvis,~G.~D.,~III; Bartlett,~R.~J. A full coupled‐cluster singles and
  doubles model: {The} inclusion of disconnected triples. \emph{J. Chem. Phys.}
  \textbf{1982}, \emph{76}, 1910--1918\relax
\mciteBstWouldAddEndPuncttrue
\mciteSetBstMidEndSepPunct{\mcitedefaultmidpunct}
{\mcitedefaultendpunct}{\mcitedefaultseppunct}\relax
\EndOfBibitem
\bibitem[Crawford and Schaefer~III(2007)Crawford, and
  Schaefer~III]{crawford_introduction_2000}
Crawford,~T.~D.; Schaefer~III,~H.~F. An introduction to coupled cluster theory
  for computational chemists. \emph{Rev. Comput. Chem.} \textbf{2007},
  \emph{14}, 33--136\relax
\mciteBstWouldAddEndPuncttrue
\mciteSetBstMidEndSepPunct{\mcitedefaultmidpunct}
{\mcitedefaultendpunct}{\mcitedefaultseppunct}\relax
\EndOfBibitem
\bibitem[Shavitt and Bartlett(2009)Shavitt, and
  Bartlett]{shavitt_many-body_2009}
Shavitt,~I.; Bartlett,~R.~J. \emph{Many-{Body} {Methods} in {Chemistry} and
  {Physics}: {MBPT} and {Coupled}-{Cluster} {Theory}}; Cambridge {Molecular}
  {Science}; Cambridge University Press: Cambridge, 2009\relax
\mciteBstWouldAddEndPuncttrue
\mciteSetBstMidEndSepPunct{\mcitedefaultmidpunct}
{\mcitedefaultendpunct}{\mcitedefaultseppunct}\relax
\EndOfBibitem
\bibitem[White(1992)]{white_density_1992}
White,~S.~R. Density matrix formulation for quantum renormalization groups.
  \emph{Phys. Rev. Lett.} \textbf{1992}, \emph{69}, 2863--2866\relax
\mciteBstWouldAddEndPuncttrue
\mciteSetBstMidEndSepPunct{\mcitedefaultmidpunct}
{\mcitedefaultendpunct}{\mcitedefaultseppunct}\relax
\EndOfBibitem
\bibitem[Chan and Sharma(2011)Chan, and Sharma]{chan_density_2011}
Chan,~G. K.-L.; Sharma,~S. The {Density} {Matrix} {Renormalization} {Group} in
  {Quantum} {Chemistry}. \emph{Annu. Rev. Phys. Chem.} \textbf{2011},
  \emph{62}, 465--481\relax
\mciteBstWouldAddEndPuncttrue
\mciteSetBstMidEndSepPunct{\mcitedefaultmidpunct}
{\mcitedefaultendpunct}{\mcitedefaultseppunct}\relax
\EndOfBibitem
\bibitem[Ceperley and Alder(1980)Ceperley, and Alder]{ceperley_ground_1980}
Ceperley,~D.~M.; Alder,~B.~J. Ground {State} of the {Electron} {Gas} by a
  {Stochastic} {Method}. \emph{Phys. Rev. Lett.} \textbf{1980}, \emph{45},
  566--569\relax
\mciteBstWouldAddEndPuncttrue
\mciteSetBstMidEndSepPunct{\mcitedefaultmidpunct}
{\mcitedefaultendpunct}{\mcitedefaultseppunct}\relax
\EndOfBibitem
\bibitem[Zhang \latin{et~al.}(1997)Zhang, Carlson, and
  Gubernatis]{zhang_constrained_1997}
Zhang,~S.; Carlson,~J.; Gubernatis,~J.~E. Constrained path {Monte} {Carlo}
  method for fermion ground states. \emph{Phys. Rev. B} \textbf{1997},
  \emph{55}, 7464--7477\relax
\mciteBstWouldAddEndPuncttrue
\mciteSetBstMidEndSepPunct{\mcitedefaultmidpunct}
{\mcitedefaultendpunct}{\mcitedefaultseppunct}\relax
\EndOfBibitem
\bibitem[Booth \latin{et~al.}(2009)Booth, Thom, and Alavi]{booth_fermion_2009}
Booth,~G.~H.; Thom,~A. J.~W.; Alavi,~A. Fermion {Monte} {Carlo} without fixed
  nodes: {A} game of life, death, and annihilation in {Slater} determinant
  space. \emph{J. Chem. Phys.} \textbf{2009}, \emph{131}, 054106\relax
\mciteBstWouldAddEndPuncttrue
\mciteSetBstMidEndSepPunct{\mcitedefaultmidpunct}
{\mcitedefaultendpunct}{\mcitedefaultseppunct}\relax
\EndOfBibitem
\bibitem[Carleo and Troyer(2017)Carleo, and Troyer]{carleo_solving_2017}
Carleo,~G.; Troyer,~M. Solving the quantum many-body problem with artificial
  neural networks. \emph{Science} \textbf{2017}, \emph{355}, 602--606\relax
\mciteBstWouldAddEndPuncttrue
\mciteSetBstMidEndSepPunct{\mcitedefaultmidpunct}
{\mcitedefaultendpunct}{\mcitedefaultseppunct}\relax
\EndOfBibitem
\bibitem[Hermann \latin{et~al.}(2020)Hermann, Schätzle, and
  Noé]{hermann_deep-neural-network_2020}
Hermann,~J.; Schätzle,~Z.; Noé,~F. Deep-neural-network solution of the
  electronic {Schrödinger} equation. \emph{Nat. Chem.} \textbf{2020},
  \emph{12}, 891--897\relax
\mciteBstWouldAddEndPuncttrue
\mciteSetBstMidEndSepPunct{\mcitedefaultmidpunct}
{\mcitedefaultendpunct}{\mcitedefaultseppunct}\relax
\EndOfBibitem
\bibitem[McMillan(1965)]{mcmillan1965ground}
McMillan,~W.~L. Ground state of liquid He 4. \emph{Phys. Rev.} \textbf{1965},
  \emph{138}, A442\relax
\mciteBstWouldAddEndPuncttrue
\mciteSetBstMidEndSepPunct{\mcitedefaultmidpunct}
{\mcitedefaultendpunct}{\mcitedefaultseppunct}\relax
\EndOfBibitem
\bibitem[Le~Roux and Bengio(2008)Le~Roux, and
  Bengio]{le_roux_representational_2008}
Le~Roux,~N.; Bengio,~Y. Representational {Power} of {Restricted} {Boltzmann}
  {Machines} and {Deep} {Belief} {Networks}. \emph{Neural Comput.}
  \textbf{2008}, \emph{20}, 1631--1649\relax
\mciteBstWouldAddEndPuncttrue
\mciteSetBstMidEndSepPunct{\mcitedefaultmidpunct}
{\mcitedefaultendpunct}{\mcitedefaultseppunct}\relax
\EndOfBibitem
\bibitem[Yang \latin{et~al.}(2020)Yang, Leng, Yu, Patel, Hu, and
  Pu]{yang_deep_2020}
Yang,~L.; Leng,~Z.; Yu,~G.; Patel,~A.; Hu,~W.-J.; Pu,~H. Deep learning-enhanced
  variational {Monte} {Carlo} method for quantum many-body physics. \emph{Phys.
  Rev. Res.} \textbf{2020}, \emph{2}, 012039\relax
\mciteBstWouldAddEndPuncttrue
\mciteSetBstMidEndSepPunct{\mcitedefaultmidpunct}
{\mcitedefaultendpunct}{\mcitedefaultseppunct}\relax
\EndOfBibitem
\bibitem[Wang \latin{et~al.}(2024)Wang, Wu, He, and Lu]{wang_variational_2024}
Wang,~J.-Q.; Wu,~H.-Q.; He,~R.-Q.; Lu,~Z.-Y. Variational optimization of the
  amplitude of neural-network quantum many-body ground states. \emph{Phys. Rev.
  B} \textbf{2024}, \emph{109}, 245120\relax
\mciteBstWouldAddEndPuncttrue
\mciteSetBstMidEndSepPunct{\mcitedefaultmidpunct}
{\mcitedefaultendpunct}{\mcitedefaultseppunct}\relax
\EndOfBibitem
\bibitem[Hibat-Allah \latin{et~al.}(2020)Hibat-Allah, Ganahl, Hayward, Melko,
  and Carrasquilla]{hibat-allah_recurrent_2020}
Hibat-Allah,~M.; Ganahl,~M.; Hayward,~L.~E.; Melko,~R.~G.; Carrasquilla,~J.
  Recurrent neural network wave functions. \emph{Phys. Rev. Res.}
  \textbf{2020}, \emph{2}, 023358\relax
\mciteBstWouldAddEndPuncttrue
\mciteSetBstMidEndSepPunct{\mcitedefaultmidpunct}
{\mcitedefaultendpunct}{\mcitedefaultseppunct}\relax
\EndOfBibitem
\bibitem[Barrett \latin{et~al.}(2022)Barrett, Malyshev, and
  Lvovsky]{barrett_autoregressive_2022}
Barrett,~T.~D.; Malyshev,~A.; Lvovsky,~A.~I. Autoregressive neural-network
  wavefunctions for ab initio quantum chemistry. \emph{Nat. Mach. Intell.}
  \textbf{2022}, \emph{4}, 351--358\relax
\mciteBstWouldAddEndPuncttrue
\mciteSetBstMidEndSepPunct{\mcitedefaultmidpunct}
{\mcitedefaultendpunct}{\mcitedefaultseppunct}\relax
\EndOfBibitem
\bibitem[Wu \latin{et~al.}(2023)Wu, Rossi, Vicentini, and
  Carleo]{wu_tensor-network_2023}
Wu,~D.; Rossi,~R.; Vicentini,~F.; Carleo,~G. From tensor-network quantum states
  to tensorial recurrent neural networks. \emph{Phys. Rev. Res.} \textbf{2023},
  \emph{5}, L032001\relax
\mciteBstWouldAddEndPuncttrue
\mciteSetBstMidEndSepPunct{\mcitedefaultmidpunct}
{\mcitedefaultendpunct}{\mcitedefaultseppunct}\relax
\EndOfBibitem
\bibitem[Wu \latin{et~al.}(2023)Wu, Guo, Fan, Zhou, and Shang]{wu2023nnqs}
Wu,~Y.; Guo,~C.; Fan,~Y.; Zhou,~P.; Shang,~H. NNQS-transformer: an efficient
  and scalable neural network quantum states approach for ab initio quantum
  chemistry. Proceedings of the International Conference for High Performance
  Computing, Networking, Storage and Analysis. 2023; pp 1--13\relax
\mciteBstWouldAddEndPuncttrue
\mciteSetBstMidEndSepPunct{\mcitedefaultmidpunct}
{\mcitedefaultendpunct}{\mcitedefaultseppunct}\relax
\EndOfBibitem
\bibitem[Viteritti \latin{et~al.}(2023)Viteritti, Rende, and
  Becca]{viteritti_transformer_2023}
Viteritti,~L.~L.; Rende,~R.; Becca,~F. Transformer {Variational} {Wave}
  {Functions} for {Frustrated} {Quantum} {Spin} {Systems}. \emph{Phys. Rev.
  Lett.} \textbf{2023}, \emph{130}, 236401\relax
\mciteBstWouldAddEndPuncttrue
\mciteSetBstMidEndSepPunct{\mcitedefaultmidpunct}
{\mcitedefaultendpunct}{\mcitedefaultseppunct}\relax
\EndOfBibitem
\bibitem[Cao \latin{et~al.}(2024)Cao, Zhong, and
  Lu]{cao2024visiontransformerneuralquantum}
Cao,~X.; Zhong,~Z.; Lu,~Y. Vision Transformer Neural Quantum States for
  Impurity Models. \emph{arXiv preprint arXiv:2408.13050} \textbf{2024}, \relax
\mciteBstWouldAddEndPunctfalse
\mciteSetBstMidEndSepPunct{\mcitedefaultmidpunct}
{}{\mcitedefaultseppunct}\relax
\EndOfBibitem
\bibitem[Hermann \latin{et~al.}(2023)Hermann, Spencer, Choo, Mezzacapo,
  Foulkes, Pfau, Carleo, and No{\'e}]{hermann2023ab}
Hermann,~J.; Spencer,~J.; Choo,~K.; Mezzacapo,~A.; Foulkes,~W. M.~C.; Pfau,~D.;
  Carleo,~G.; No{\'e},~F. Ab initio quantum chemistry with neural-network
  wavefunctions. \emph{Nature Reviews Chemistry} \textbf{2023}, \emph{7},
  692--709\relax
\mciteBstWouldAddEndPuncttrue
\mciteSetBstMidEndSepPunct{\mcitedefaultmidpunct}
{\mcitedefaultendpunct}{\mcitedefaultseppunct}\relax
\EndOfBibitem
\bibitem[Levin and Peres(2017)Levin, and Peres]{levin_markov_2017}
Levin,~D.; Peres,~Y. \emph{Markov {Chains} and {Mixing} {Times}}; American
  Mathematical Society: Providence, Rhode Island, 2017\relax
\mciteBstWouldAddEndPuncttrue
\mciteSetBstMidEndSepPunct{\mcitedefaultmidpunct}
{\mcitedefaultendpunct}{\mcitedefaultseppunct}\relax
\EndOfBibitem
\bibitem[Wolff(1990)]{wolff1990critical}
Wolff,~U. Critical slowing down. \emph{Nucl. Phys. B Proc. Suppl.}
  \textbf{1990}, \emph{17}, 93--102\relax
\mciteBstWouldAddEndPuncttrue
\mciteSetBstMidEndSepPunct{\mcitedefaultmidpunct}
{\mcitedefaultendpunct}{\mcitedefaultseppunct}\relax
\EndOfBibitem
\bibitem[Jiang \latin{et~al.}(2024)Jiang, Zhang, Baumgarten, Chen, Dinh,
  Ganeshram, Maskara, Ni, and Lee]{jiang2024walkinghilbertspacequantum}
Jiang,~T.; Zhang,~J.; Baumgarten,~M.~K.; Chen,~M.-F.; Dinh,~H.~Q.;
  Ganeshram,~A.; Maskara,~N.; Ni,~A.; Lee,~J. Walking through Hilbert space
  with quantum computers. \emph{arXiv preprint arXiv:2407.11672} \textbf{2024},
  \relax
\mciteBstWouldAddEndPunctfalse
\mciteSetBstMidEndSepPunct{\mcitedefaultmidpunct}
{}{\mcitedefaultseppunct}\relax
\EndOfBibitem
\bibitem[Arute \latin{et~al.}(2019)Arute, Arya, Babbush, Bacon, Bardin,
  Barends, Biswas, Boixo, Brandao, Buell, Burkett, Chen, Chen, Chiaro, Collins,
  Courtney, Dunsworth, Farhi, Foxen, Fowler, Gidney, Giustina, Graff, Guerin,
  Habegger, Harrigan, Hartmann, Ho, Hoffmann, Huang, Humble, Isakov, Jeffrey,
  Jiang, Kafri, Kechedzhi, Kelly, Klimov, Knysh, Korotkov, Kostritsa, Landhuis,
  Lindmark, Lucero, Lyakh, Mandrà, McClean, McEwen, Megrant, Mi, Michielsen,
  Mohseni, Mutus, Naaman, Neeley, Neill, Niu, Ostby, Petukhov, Platt, Quintana,
  Rieffel, Roushan, Rubin, Sank, Satzinger, Smelyanskiy, Sung, Trevithick,
  Vainsencher, Villalonga, White, Yao, Yeh, Zalcman, Neven, and
  Martinis]{arute_quantum_2019}
Arute,~F. \latin{et~al.}  Quantum supremacy using a programmable
  superconducting processor. \emph{Nature} \textbf{2019}, \emph{574},
  505--510\relax
\mciteBstWouldAddEndPuncttrue
\mciteSetBstMidEndSepPunct{\mcitedefaultmidpunct}
{\mcitedefaultendpunct}{\mcitedefaultseppunct}\relax
\EndOfBibitem
\bibitem[Wu \latin{et~al.}(2021)Wu, Bao, Cao, Chen, Chen, Chen, Chung, Deng,
  Du, Fan, Gong, Guo, Guo, Guo, Han, Hong, Huang, Huo, Li, Li, Li, Li, Liang,
  Lin, Lin, Qian, Qiao, Rong, Su, Sun, Wang, Wang, Wu, Xu, Yan, Yang, Yang, Ye,
  Yin, Ying, Yu, Zha, Zhang, Zhang, Zhang, Zhang, Zhao, Zhao, Zhou, Zhu, Lu,
  Peng, Zhu, and Pan]{wu_strong_2021}
Wu,~Y. \latin{et~al.}  Strong {Quantum} {Computational} {Advantage} {Using} a
  {Superconducting} {Quantum} {Processor}. \emph{Phys. Rev. Lett.}
  \textbf{2021}, \emph{127}, 180501\relax
\mciteBstWouldAddEndPuncttrue
\mciteSetBstMidEndSepPunct{\mcitedefaultmidpunct}
{\mcitedefaultendpunct}{\mcitedefaultseppunct}\relax
\EndOfBibitem
\bibitem[Cao \latin{et~al.}(2019)Cao, Romero, Olson, Degroote, Johnson,
  Kieferová, Kivlichan, Menke, Peropadre, Sawaya, Sim, Veis, and
  Aspuru-Guzik]{cao_quantum_2019}
Cao,~Y.; Romero,~J.; Olson,~J.~P.; Degroote,~M.; Johnson,~P.~D.;
  Kieferová,~M.; Kivlichan,~I.~D.; Menke,~T.; Peropadre,~B.; Sawaya,~N. P.~D.;
  Sim,~S.; Veis,~L.; Aspuru-Guzik,~A. Quantum {Chemistry} in the {Age} of
  {Quantum} {Computing}. \emph{Chem. Rev.} \textbf{2019}, \emph{119},
  10856--10915\relax
\mciteBstWouldAddEndPuncttrue
\mciteSetBstMidEndSepPunct{\mcitedefaultmidpunct}
{\mcitedefaultendpunct}{\mcitedefaultseppunct}\relax
\EndOfBibitem
\bibitem[McArdle \latin{et~al.}(2020)McArdle, Endo, Aspuru-Guzik, Benjamin, and
  Yuan]{mcardle_quantum_2020}
McArdle,~S.; Endo,~S.; Aspuru-Guzik,~A.; Benjamin,~S.~C.; Yuan,~X. Quantum
  computational chemistry. \emph{Rev. Mod. Phys.} \textbf{2020}, \emph{92},
  015003\relax
\mciteBstWouldAddEndPuncttrue
\mciteSetBstMidEndSepPunct{\mcitedefaultmidpunct}
{\mcitedefaultendpunct}{\mcitedefaultseppunct}\relax
\EndOfBibitem
\bibitem[Bauer \latin{et~al.}(2020)Bauer, Bravyi, Motta, and
  Chan]{bauer_quantum_2020}
Bauer,~B.; Bravyi,~S.; Motta,~M.; Chan,~G. K.-L. Quantum {Algorithms} for
  {Quantum} {Chemistry} and {Quantum} {Materials} {Science}. \emph{Chem. Rev.}
  \textbf{2020}, \emph{120}, 12685--12717\relax
\mciteBstWouldAddEndPuncttrue
\mciteSetBstMidEndSepPunct{\mcitedefaultmidpunct}
{\mcitedefaultendpunct}{\mcitedefaultseppunct}\relax
\EndOfBibitem
\bibitem[Motta and Rice(2022)Motta, and Rice]{motta_emerging_2022}
Motta,~M.; Rice,~J.~E. Emerging quantum computing algorithms for quantum
  chemistry. \emph{WIREs Comput. Mol. Sci.} \textbf{2022}, \emph{12},
  e1580\relax
\mciteBstWouldAddEndPuncttrue
\mciteSetBstMidEndSepPunct{\mcitedefaultmidpunct}
{\mcitedefaultendpunct}{\mcitedefaultseppunct}\relax
\EndOfBibitem
\bibitem[Szegedy(2004)]{szegedy_quantum_2004}
Szegedy,~M. Quantum speed-up of {Markov} chain based algorithms. 45th {Annual}
  {IEEE} {Symposium} on {Foundations} of {Computer} {Science}. 2004; pp
  32--41\relax
\mciteBstWouldAddEndPuncttrue
\mciteSetBstMidEndSepPunct{\mcitedefaultmidpunct}
{\mcitedefaultendpunct}{\mcitedefaultseppunct}\relax
\EndOfBibitem
\bibitem[Somma \latin{et~al.}(2008)Somma, Boixo, Barnum, and
  Knill]{somma_quantum_2008}
Somma,~R.~D.; Boixo,~S.; Barnum,~H.; Knill,~E. Quantum {Simulations} of
  {Classical} {Annealing} {Processes}. \emph{Phys. Rev. Lett.} \textbf{2008},
  \emph{101}, 130504\relax
\mciteBstWouldAddEndPuncttrue
\mciteSetBstMidEndSepPunct{\mcitedefaultmidpunct}
{\mcitedefaultendpunct}{\mcitedefaultseppunct}\relax
\EndOfBibitem
\bibitem[Wocjan and Abeyesinghe(2008)Wocjan, and
  Abeyesinghe]{wocjan_speedup_2008}
Wocjan,~P.; Abeyesinghe,~A. Speedup via quantum sampling. \emph{Phys. Rev. A}
  \textbf{2008}, \emph{78}, 042336\relax
\mciteBstWouldAddEndPuncttrue
\mciteSetBstMidEndSepPunct{\mcitedefaultmidpunct}
{\mcitedefaultendpunct}{\mcitedefaultseppunct}\relax
\EndOfBibitem
\bibitem[Poulin and Wocjan(2009)Poulin, and Wocjan]{poulin_sampling_2009}
Poulin,~D.; Wocjan,~P. Sampling from the {Thermal} {Quantum} {Gibbs} {State}
  and {Evaluating} {Partition} {Functions} with a {Quantum} {Computer}.
  \emph{Phys. Rev. Lett.} \textbf{2009}, \emph{103}, 220502\relax
\mciteBstWouldAddEndPuncttrue
\mciteSetBstMidEndSepPunct{\mcitedefaultmidpunct}
{\mcitedefaultendpunct}{\mcitedefaultseppunct}\relax
\EndOfBibitem
\bibitem[Bilgin and Boixo(2010)Bilgin, and Boixo]{bilgin_preparing_2010}
Bilgin,~E.; Boixo,~S. Preparing {Thermal} {States} of {Quantum} {Systems} by
  {Dimension} {Reduction}. \emph{Phys. Rev. Lett.} \textbf{2010}, \emph{105},
  170405\relax
\mciteBstWouldAddEndPuncttrue
\mciteSetBstMidEndSepPunct{\mcitedefaultmidpunct}
{\mcitedefaultendpunct}{\mcitedefaultseppunct}\relax
\EndOfBibitem
\bibitem[Temme \latin{et~al.}(2011)Temme, Osborne, Vollbrecht, Poulin, and
  Verstraete]{temme_quantum_2011}
Temme,~K.; Osborne,~T.~J.; Vollbrecht,~K.~G.; Poulin,~D.; Verstraete,~F.
  Quantum {Metropolis} sampling. \emph{Nature} \textbf{2011}, \emph{471},
  87--90\relax
\mciteBstWouldAddEndPuncttrue
\mciteSetBstMidEndSepPunct{\mcitedefaultmidpunct}
{\mcitedefaultendpunct}{\mcitedefaultseppunct}\relax
\EndOfBibitem
\bibitem[Yung and Aspuru-Guzik(2012)Yung, and
  Aspuru-Guzik]{yung_quantumquantum_2012}
Yung,~M.-H.; Aspuru-Guzik,~A. A quantum–quantum {Metropolis} algorithm.
  \emph{Proc. Natl Acad. Sci. USA} \textbf{2012}, \emph{109}, 754--759\relax
\mciteBstWouldAddEndPuncttrue
\mciteSetBstMidEndSepPunct{\mcitedefaultmidpunct}
{\mcitedefaultendpunct}{\mcitedefaultseppunct}\relax
\EndOfBibitem
\bibitem[Montanaro(2015)]{montanaro_quantum_2015}
Montanaro,~A. Quantum speedup of {Monte} {Carlo} methods. \emph{Proc. R. Soc.
  A} \textbf{2015}, \emph{471}, 20150301\relax
\mciteBstWouldAddEndPuncttrue
\mciteSetBstMidEndSepPunct{\mcitedefaultmidpunct}
{\mcitedefaultendpunct}{\mcitedefaultseppunct}\relax
\EndOfBibitem
\bibitem[Chowdhury and Somma(2017)Chowdhury, and Somma]{chowdhury_quantum_2017}
Chowdhury,~A.~N.; Somma,~R.~D. Quantum algorithms for {Gibbs} sampling and
  hitting-time estimation. \emph{Quantum Inf. Comput.} \textbf{2017},
  \emph{17}, 41--64\relax
\mciteBstWouldAddEndPuncttrue
\mciteSetBstMidEndSepPunct{\mcitedefaultmidpunct}
{\mcitedefaultendpunct}{\mcitedefaultseppunct}\relax
\EndOfBibitem
\bibitem[Lemieux \latin{et~al.}(2020)Lemieux, Heim, Poulin, Svore, and
  Troyer]{lemieux_efficient_2020}
Lemieux,~J.; Heim,~B.; Poulin,~D.; Svore,~K.; Troyer,~M. Efficient {Quantum}
  {Walk} {Circuits} for {Metropolis}-{Hastings} {Algorithm}. \emph{Quantum}
  \textbf{2020}, \emph{4}, 287\relax
\mciteBstWouldAddEndPuncttrue
\mciteSetBstMidEndSepPunct{\mcitedefaultmidpunct}
{\mcitedefaultendpunct}{\mcitedefaultseppunct}\relax
\EndOfBibitem
\bibitem[Arunachalam \latin{et~al.}(2022)Arunachalam, Havlicek, Nannicini,
  Temme, and Wocjan]{arunachalam_simpler_2022}
Arunachalam,~S.; Havlicek,~V.; Nannicini,~G.; Temme,~K.; Wocjan,~P. Simpler
  (classical) and faster (quantum) algorithms for {Gibbs} partition functions.
  \emph{Quantum} \textbf{2022}, \emph{6}, 789\relax
\mciteBstWouldAddEndPuncttrue
\mciteSetBstMidEndSepPunct{\mcitedefaultmidpunct}
{\mcitedefaultendpunct}{\mcitedefaultseppunct}\relax
\EndOfBibitem
\bibitem[Rall \latin{et~al.}(2023)Rall, Wang, and Wocjan]{rall_thermal_2023}
Rall,~P.; Wang,~C.; Wocjan,~P. Thermal {State} {Preparation} via {Rounding}
  {Promises}. \emph{Quantum} \textbf{2023}, \emph{7}, 1132\relax
\mciteBstWouldAddEndPuncttrue
\mciteSetBstMidEndSepPunct{\mcitedefaultmidpunct}
{\mcitedefaultendpunct}{\mcitedefaultseppunct}\relax
\EndOfBibitem
\bibitem[Chen \latin{et~al.}(2023)Chen, Kastoryano, and
  Gily{\'e}n]{chen_efficient_2023}
Chen,~C.-F.; Kastoryano,~M.~J.; Gily{\'e}n,~A. An efficient and exact
  noncommutative quantum Gibbs sampler. \emph{arXiv preprint arXiv:2311.09207}
  \textbf{2023}, \relax
\mciteBstWouldAddEndPunctfalse
\mciteSetBstMidEndSepPunct{\mcitedefaultmidpunct}
{}{\mcitedefaultseppunct}\relax
\EndOfBibitem
\bibitem[Wild \latin{et~al.}(2021)Wild, Sels, Pichler, Zanoci, and
  Lukin]{wild_quantum_2021}
Wild,~D.~S.; Sels,~D.; Pichler,~H.; Zanoci,~C.; Lukin,~M.~D. Quantum {Sampling}
  {Algorithms} for {Near}-{Term} {Devices}. \emph{Phys. Rev. Lett.}
  \textbf{2021}, \emph{127}, 100504\relax
\mciteBstWouldAddEndPuncttrue
\mciteSetBstMidEndSepPunct{\mcitedefaultmidpunct}
{\mcitedefaultendpunct}{\mcitedefaultseppunct}\relax
\EndOfBibitem
\bibitem[Wild \latin{et~al.}(2021)Wild, Sels, Pichler, Zanoci, and
  Lukin]{wild_quantum_2021-1}
Wild,~D.~S.; Sels,~D.; Pichler,~H.; Zanoci,~C.; Lukin,~M.~D. Quantum sampling
  algorithms, phase transitions, and computational complexity. \emph{Phys. Rev.
  A} \textbf{2021}, \emph{104}, 032602\relax
\mciteBstWouldAddEndPuncttrue
\mciteSetBstMidEndSepPunct{\mcitedefaultmidpunct}
{\mcitedefaultendpunct}{\mcitedefaultseppunct}\relax
\EndOfBibitem
\bibitem[Layden \latin{et~al.}(2023)Layden, Mazzola, Mishmash, Motta, Wocjan,
  Kim, and Sheldon]{layden_quantum-enhanced_2023}
Layden,~D.; Mazzola,~G.; Mishmash,~R.~V.; Motta,~M.; Wocjan,~P.; Kim,~J.-S.;
  Sheldon,~S. Quantum-enhanced {Markov} chain {Monte} {Carlo}. \emph{Nature}
  \textbf{2023}, \emph{619}, 282--287\relax
\mciteBstWouldAddEndPuncttrue
\mciteSetBstMidEndSepPunct{\mcitedefaultmidpunct}
{\mcitedefaultendpunct}{\mcitedefaultseppunct}\relax
\EndOfBibitem
\bibitem[Nakano \latin{et~al.}(2024)Nakano, Hakoshima, Mitarai, and
  Fujii]{nakano_markov-chain_2024}
Nakano,~Y.; Hakoshima,~H.; Mitarai,~K.; Fujii,~K. Markov-chain {Monte} {Carlo}
  method enhanced by a quantum alternating operator ansatz. \emph{Phys. Rev.
  Res.} \textbf{2024}, \emph{6}, 033105\relax
\mciteBstWouldAddEndPuncttrue
\mciteSetBstMidEndSepPunct{\mcitedefaultmidpunct}
{\mcitedefaultendpunct}{\mcitedefaultseppunct}\relax
\EndOfBibitem
\bibitem[Ding \latin{et~al.}(2024)Ding, Chen, and
  Lin]{PhysRevResearch.6.033147}
Ding,~Z.; Chen,~C.-F.; Lin,~L. Single-ancilla ground state preparation via
  Lindbladians. \emph{Phys. Rev. Res.} \textbf{2024}, \emph{6}, 033147\relax
\mciteBstWouldAddEndPuncttrue
\mciteSetBstMidEndSepPunct{\mcitedefaultmidpunct}
{\mcitedefaultendpunct}{\mcitedefaultseppunct}\relax
\EndOfBibitem
\bibitem[Orfi and Sels(2024)Orfi, and Sels]{orfi_bounding_2024}
Orfi,~A.; Sels,~D. Bounding the speedup of the quantum-enhanced {Markov}-chain
  {Monte} {Carlo} algorithm. \emph{Phys. Rev. A} \textbf{2024}, \emph{110},
  052414\relax
\mciteBstWouldAddEndPuncttrue
\mciteSetBstMidEndSepPunct{\mcitedefaultmidpunct}
{\mcitedefaultendpunct}{\mcitedefaultseppunct}\relax
\EndOfBibitem
\bibitem[Orfi and Sels(2024)Orfi, and Sels]{orfi_barriers_2024}
Orfi,~A.; Sels,~D. Barriers to efficient mixing of quantum-enhanced {Markov}
  chains. \emph{Phys. Rev. A} \textbf{2024}, \emph{110}, 052434\relax
\mciteBstWouldAddEndPuncttrue
\mciteSetBstMidEndSepPunct{\mcitedefaultmidpunct}
{\mcitedefaultendpunct}{\mcitedefaultseppunct}\relax
\EndOfBibitem
\bibitem[Christmann \latin{et~al.}(2024)Christmann, Ivashkov, Chiurco, and
  Mazzola]{christmann_quantum_2024}
Christmann,~J.; Ivashkov,~P.; Chiurco,~M.; Mazzola,~G. From quantum enhanced to
  quantum inspired Monte Carlo. \emph{arXiv preprint arXiv:2411.17821}
  \textbf{2024}, \relax
\mciteBstWouldAddEndPunctfalse
\mciteSetBstMidEndSepPunct{\mcitedefaultmidpunct}
{}{\mcitedefaultseppunct}\relax
\EndOfBibitem
\bibitem[Lockwood \latin{et~al.}(2024)Lockwood, Weiss, Aronshtein, and
  Verdon]{lockwood_quantum_2024}
Lockwood,~O.; Weiss,~P.; Aronshtein,~F.; Verdon,~G. Quantum dynamical
  {Hamiltonian} {Monte} {Carlo}. \emph{Phys. Rev. Res.} \textbf{2024},
  \emph{6}, 033142\relax
\mciteBstWouldAddEndPuncttrue
\mciteSetBstMidEndSepPunct{\mcitedefaultmidpunct}
{\mcitedefaultendpunct}{\mcitedefaultseppunct}\relax
\EndOfBibitem
\bibitem[Ferguson and Wallden(2024)Ferguson, and
  Wallden]{ferguson_quantum-enhanced_2024}
Ferguson,~S.; Wallden,~P. Quantum-enhanced Markov Chain Monte Carlo for systems
  larger than your Quantum Computer. \emph{arXiv preprint arXiv:2405.04247}
  \textbf{2024}, \relax
\mciteBstWouldAddEndPunctfalse
\mciteSetBstMidEndSepPunct{\mcitedefaultmidpunct}
{}{\mcitedefaultseppunct}\relax
\EndOfBibitem
\bibitem[Sajjan \latin{et~al.}(2024)Sajjan, Singh, and
  Kais]{sajjan2024polynomiallyefficientquantumenabled}
Sajjan,~M.; Singh,~V.; Kais,~S. Polynomially efficient quantum enabled
  variational Monte Carlo for training neural-network quantum states for
  physico-chemical applications. \emph{arXiv preprint arXiv:2412.12398}
  \textbf{2024}, \relax
\mciteBstWouldAddEndPunctfalse
\mciteSetBstMidEndSepPunct{\mcitedefaultmidpunct}
{}{\mcitedefaultseppunct}\relax
\EndOfBibitem
\bibitem[Sorella(2005)]{sorella_wave_2005}
Sorella,~S. Wave function optimization in the variational {Monte} {Carlo}
  method. \emph{Phys. Rev. B} \textbf{2005}, \emph{71}, 241103\relax
\mciteBstWouldAddEndPuncttrue
\mciteSetBstMidEndSepPunct{\mcitedefaultmidpunct}
{\mcitedefaultendpunct}{\mcitedefaultseppunct}\relax
\EndOfBibitem
\bibitem[Pfau \latin{et~al.}(2020)Pfau, Spencer, Matthews, and
  Foulkes]{pfau_ab_2020}
Pfau,~D.; Spencer,~J.~S.; Matthews,~A. G. D.~G.; Foulkes,~W. M.~C. Ab initio
  solution of the many-electron Schr\"odinger equation with deep neural
  networks. \emph{Phys. Rev. Res.} \textbf{2020}, \emph{2}, 033429\relax
\mciteBstWouldAddEndPuncttrue
\mciteSetBstMidEndSepPunct{\mcitedefaultmidpunct}
{\mcitedefaultendpunct}{\mcitedefaultseppunct}\relax
\EndOfBibitem
\bibitem[Choo \latin{et~al.}(2020)Choo, Mezzacapo, and
  Carleo]{choo_fermionic_2020}
Choo,~K.; Mezzacapo,~A.; Carleo,~G. Fermionic neural-network states for
  ab-initio electronic structure. \emph{Nat. Commun.} \textbf{2020}, \emph{11},
  2368\relax
\mciteBstWouldAddEndPuncttrue
\mciteSetBstMidEndSepPunct{\mcitedefaultmidpunct}
{\mcitedefaultendpunct}{\mcitedefaultseppunct}\relax
\EndOfBibitem
\bibitem[Torlai \latin{et~al.}(2018)Torlai, Mazzola, Carrasquilla, Troyer,
  Melko, and Carleo]{torlai_neural-network_2018}
Torlai,~G.; Mazzola,~G.; Carrasquilla,~J.; Troyer,~M.; Melko,~R.; Carleo,~G.
  Neural-network quantum state tomography. \emph{Nat. Phys.} \textbf{2018},
  \emph{14}, 447--450\relax
\mciteBstWouldAddEndPuncttrue
\mciteSetBstMidEndSepPunct{\mcitedefaultmidpunct}
{\mcitedefaultendpunct}{\mcitedefaultseppunct}\relax
\EndOfBibitem
\bibitem[Kingma(2014)]{kingma2017adammethodstochasticoptimization}
Kingma,~D.~P. Adam: A method for stochastic optimization. \emph{arXiv preprint
  arXiv:1412.6980} \textbf{2014}, \relax
\mciteBstWouldAddEndPunctfalse
\mciteSetBstMidEndSepPunct{\mcitedefaultmidpunct}
{}{\mcitedefaultseppunct}\relax
\EndOfBibitem
\bibitem[Metropolis \latin{et~al.}(1953)Metropolis, Rosenbluth, Rosenbluth,
  Teller, and Teller]{metropolis_equation_1953}
Metropolis,~N.; Rosenbluth,~A.~W.; Rosenbluth,~M.~N.; Teller,~A.~H.; Teller,~E.
  Equation of {State} {Calculations} by {Fast} {Computing} {Machines}. \emph{J.
  Chem. Phys.} \textbf{1953}, \emph{21}, 1087--1092\relax
\mciteBstWouldAddEndPuncttrue
\mciteSetBstMidEndSepPunct{\mcitedefaultmidpunct}
{\mcitedefaultendpunct}{\mcitedefaultseppunct}\relax
\EndOfBibitem
\bibitem[cha()]{changQAVMC2024}
Quantum-assisted variational Monte Carlo.
  \url{https://github.com/dilandaer/QA-VMC}\relax
\mciteBstWouldAddEndPuncttrue
\mciteSetBstMidEndSepPunct{\mcitedefaultmidpunct}
{\mcitedefaultendpunct}{\mcitedefaultseppunct}\relax
\EndOfBibitem
\bibitem[Sokal(1997)]{Sokal1997}
Sokal,~A. In \emph{Functional Integration: Basics and Applications};
  DeWitt-Morette,~C., Cartier,~P., Folacci,~A., Eds.; Springer US: Boston, MA,
  1997; pp 131--192\relax
\mciteBstWouldAddEndPuncttrue
\mciteSetBstMidEndSepPunct{\mcitedefaultmidpunct}
{\mcitedefaultendpunct}{\mcitedefaultseppunct}\relax
\EndOfBibitem
\bibitem[Arovas \latin{et~al.}(2022)Arovas, Berg, Kivelson, and
  Raghu]{arovas_hubbard_2022}
Arovas,~D.~P.; Berg,~E.; Kivelson,~S.~A.; Raghu,~S. The {Hubbard} {Model}.
  \emph{Annu. Rev. Condens. Matter Phys.} \textbf{2022}, \emph{13},
  239--274\relax
\mciteBstWouldAddEndPuncttrue
\mciteSetBstMidEndSepPunct{\mcitedefaultmidpunct}
{\mcitedefaultendpunct}{\mcitedefaultseppunct}\relax
\EndOfBibitem
\bibitem[Yokoyama and Shiba(1987)Yokoyama, and
  Shiba]{yokoyama_variational_1987}
Yokoyama,~H.; Shiba,~H. Variational {Monte}-{Carlo} {Studies} of {Hubbard}
  {Model}. {I}. \emph{J. Phys. Soc. Jpn.} \textbf{1987}, \emph{56},
  1490--1506\relax
\mciteBstWouldAddEndPuncttrue
\mciteSetBstMidEndSepPunct{\mcitedefaultmidpunct}
{\mcitedefaultendpunct}{\mcitedefaultseppunct}\relax
\EndOfBibitem
\bibitem[Cade \latin{et~al.}(2020)Cade, Mineh, Montanaro, and
  Stanisic]{cade_strategies_2020}
Cade,~C.; Mineh,~L.; Montanaro,~A.; Stanisic,~S. Strategies for solving the
  {Fermi}-{Hubbard} model on near-term quantum computers. \emph{Phys. Rev. B}
  \textbf{2020}, \emph{102}, 235122\relax
\mciteBstWouldAddEndPuncttrue
\mciteSetBstMidEndSepPunct{\mcitedefaultmidpunct}
{\mcitedefaultendpunct}{\mcitedefaultseppunct}\relax
\EndOfBibitem
\bibitem[Jordan and Wigner(1928)Jordan, and Wigner]{jordan1928pauli}
Jordan,~P.; Wigner,~E.~P. About the Pauli exclusion principle. \emph{Z. Phys}
  \textbf{1928}, \emph{47}, 14--75\relax
\mciteBstWouldAddEndPuncttrue
\mciteSetBstMidEndSepPunct{\mcitedefaultmidpunct}
{\mcitedefaultendpunct}{\mcitedefaultseppunct}\relax
\EndOfBibitem
\end{mcitethebibliography}

\end{document}